\documentclass[journal]{IEEEtran}
\usepackage{amsmath,amsfonts}
\usepackage{algorithmic}
\usepackage{algorithm}
\usepackage{array}
\usepackage[caption=false,font=normalsize,labelfont=sf,textfont=sf]{subfig}
\usepackage{textcomp}
\usepackage{stfloats}
\usepackage{url}
\usepackage{verbatim}
\usepackage{graphicx}
\usepackage{multirow}
\usepackage{xcolor}
\DeclareRobustCommand\blue{\textcolor{black}}
\usepackage[square,numbers]{natbib}
\usepackage{multicol}
\usepackage{capt-of}
\usepackage{pifont}
\newcommand{\cmark}{\ding{51}} 
\newcommand{\xmark}{\ding{55}} 
\newcolumntype{K}[1]{>{\centering\arraybackslash}p{#1}}

\begin{document}	
	\title{ Design and Optimization of a Hybrid VLC/THz Infrastructure-to-Vehicle Communication System for Intelligent  Transportation}
	\author{Yusef Modami, Hamzeh Beiranvand, and Mohammad Taghi Dabiri
 \vspace{-0.15cm}
		\thanks{Yusef Modami and Hamzeh Beyranvad are with the Department of Electrical Engineering, Amirkabir University of Technology (Tehran Polytechnic), Tehran, Iran (email: y.modami@aut.ac.ir; beyranvand@aut.ac.ir). Mohammad T. Dabiri is with the Department of Electrical Engineering, Qatar University, Doha, Qatar (e-mail: m.dabiri@qu.edu.qa)}}  
	

	
	\maketitle

	\begin{abstract} 
  \textcolor{black}{This paper proposes a hybrid infrastructure-to-vehicle (I2V) communication framework to support future 6G-enabled intelligent transportation systems (ITS) in smart cities. Leveraging existing LED streetlighting infrastructure, the system simultaneously delivers energy-efficient illumination and high-speed wireless connectivity. The proposed scheme integrates visible light communication (VLC) with a complementary terahertz (THz) antenna array to overcome VLC limitations under high ambient light and adverse weather conditions. Key contributions include the design of a VLC/THz access network, seamless integration with lighting infrastructure, a proposed switching-combination (PSC) mechanism, and a physical layout optimization strategy. Using a grid search method, thousands of configurations were evaluated to maximize lighting coverage, received power, signal-to-noise ratio (SNR), signal-to-interference-and-noise ratio (SINR), and minimize outage probability. Results show that optimized lighting coverage improves from 35\% to 97\%, while hybrid communication coverage increases from 49\% to 99.9\% at the same power level. Under extreme environmental conditions, the hybrid system maintains up to 99\% coverage, compared to 69\% with VLC alone. These results demonstrate the scalability, cost-efficiency, and practicality of the proposed system for next-generation ITS deployment.}
  \end{abstract}

\vspace{-0.002cm}

\section{Introduction}
\begin{figure}[t]
		\centering
		\includegraphics[width=0.8\linewidth,height=0.14\textheight]{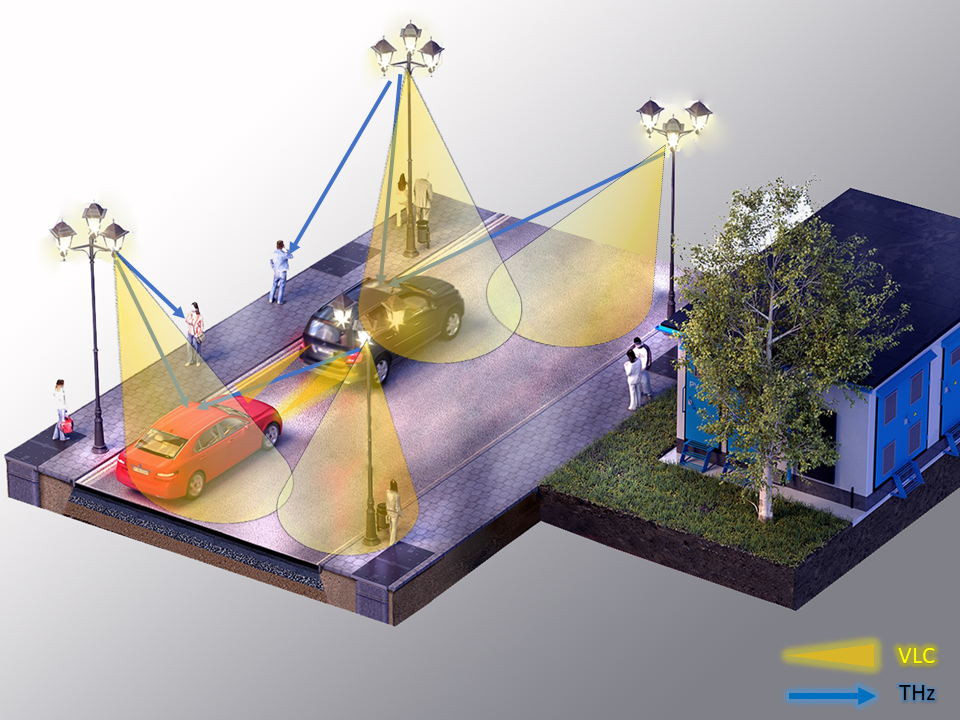}
		\vspace{-0.2cm}
		\caption{Illustration of hybrid VLC/THz-based I2V communication.}
		\vspace{-0.6cm}
		\label{fig2}
	\end{figure}

 \IEEEPARstart{\blue{S}}{}
  \hspace{-0.2cm}\blue{mart city dream, powered by 6G networks and dense internet of things (IoT) deployments, is steadily transitioning from concept to reality. These urban ecosystems promise significant improvements in efficiency, sustainability, and quality of life \cite{ref3,ref4}. Within this context, ITS is a foundational component, integrating vehicles, pedestrians, and infrastructure into a cohesive network that enables vehicle-to-vehicle (V2V), infrastructure-to-vehicle (I2V), and infrastructure-to-infrastructure (I2I) communications (Fig.~\ref{fig2}). Such integration is essential for enhancing road safety, optimizing traffic flow, and supporting 6G-connected autonomous vehicles  \cite{ref28,ref29}.}
    
	\textcolor{black}{However, the extreme data-rate and ultra-low-latency demands of next-generation ITS—driven by massive sensor deployments and real-time control—far exceed the capabilities of existing radio frequency (RF) networks \cite{ref30}. As a result, research has increasingly focused on higher-frequency communication technologies, including millimeter-wave (mmWave), terahertz (THz), and optical bands. While these bands offer abundant bandwidth, they also pose significant challenges \cite{ref12}. This raises a critical question: which technology, or combination of technologies, can reliably meet the stringent performance requirements of future ITS?}

	\textcolor{black}{The growing deployment of light-emitting diodes (LEDs) in urban streetlights and vehicles offers a promising answer. These LED infrastructures are well-suited for VLC implementation, which leverages the rapid switching capabilities of LEDs to achieve data rates exceeding 1 Gbps, all without adverse health effects \cite{ref19}. Furthermore, VLC operates in the unlicensed optical spectrum, offering far greater bandwidth than traditional RF systems \cite{ref21,ref22}. By enabling simultaneous lighting and high-speed data transmission, VLC stands out as a promising solution for smart city wireless infrastructure and ITS applications \cite{ref13,ref13_1}.}

	\textcolor{black}{Despite its advantages, outdoor VLC faces two key limitations. First, ambient sunlight and artificial lighting introduce significant background noise, degrading the SNR. Second, adverse weather conditions, such as rain, fog, snow, and airborne particulates, scatter and attenuate optical signals, leading to performance degradation and potential service interruptions. Additionally, maintaining communication when streetlights are turned off poses a practical challenge. One approach involves dimming LEDs to a level that appears off to the human eye while still enabling data transmission \cite{ref48}. A more robust solution, however, is to adopt a hybrid communication system that addresses VLC’s inherent limitations.}

 \begin{table*}\vspace{-0.6cm}
\caption{\textcolor{black}{Comparison of This Study with Related Works}}
 \vspace{-0.15cm}
\begin{minipage}{0.99\textwidth}
\centering
\renewcommand{\arraystretch}{1.3}
\textcolor{black}{\begin{tabular}{|l|K{0.47cm}|K{0.47cm}|K{0.47cm}|K{0.47cm}|K{0.47cm}|K{0.47cm}|K{0.47cm}|K{0.47cm}|K{0.47cm}|K{0.47cm}|K{0.47cm}|K{0.47cm}|c|}
\hline
\textbf{Aspect} & \cite{ref29} & \cite{ref114} & \cite{ref115} & \cite{ref77} & \cite{ref73} & \cite{ref83} & \cite{ref49} & \cite{ref62} & \cite{ref56} & \cite{ref74} & \cite{ref75} & \cite{ref76} &\!\!\!This Study\!\!\!\hspace{-0.1cm}\\
\hline
VLC-based I2V communication & \cmark & \xmark & \xmark & \cmark & \cmark & \cmark & \xmark & \xmark & \xmark & \xmark &\cmark & \cmark & \cmark \\
\hline
Integration of THz with VLC & \xmark & \cmark & \xmark & \xmark & \xmark & \xmark & \xmark & \xmark & \xmark & \xmark &\xmark & \xmark & \cmark \\
\hline
Hybrid communication systems  & \xmark & \cmark & \cmark & \xmark & \xmark & \xmark & \cmark & \cmark & \cmark & \xmark &\xmark & \xmark & \cmark \\
\hline
Physical components design & \xmark & \xmark & \xmark & \xmark & \cmark & \xmark & \xmark & \xmark & \xmark & \cmark &\cmark & \xmark & \cmark \\
\hline
Integration of VLC with lighting infrastructure & \cmark & \xmark & \xmark & \cmark & \cmark & \cmark & \xmark & \xmark & \xmark & \xmark &\cmark & \xmark & \cmark \\
\hline
Mobility, handover, and latency handling & \xmark & \xmark & \cmark & \cmark & \xmark & \xmark & \cmark & \xmark & \xmark & \xmark & \xmark & \cmark & \cmark \\
\hline
System parameter optimization & \xmark & \xmark & \xmark & \xmark & \cmark & \cmark & \xmark & \xmark & \xmark & \cmark & \cmark & \xmark & \cmark \\
\hline
Switching/link adaptation mechanism & \xmark & \xmark & \xmark & \xmark & \xmark & \xmark & \xmark & \cmark & \cmark & \xmark & \xmark & \xmark & \cmark \\
\hline
Fully end-to-end integrated I2V architecture  & \xmark & \xmark & \xmark & \xmark & \xmark & \xmark & \xmark & \xmark & \xmark & \xmark & \xmark & \xmark & \cmark \\
\hline
\end{tabular}}
			\label{jad0}
	\end{minipage}
 \vspace{-0.45cm}
 \end{table*}
    
	\textcolor{black}{ The THz band (100 GHz to 10 THz) offers extremely high bandwidth and inherent security, making it ideal for high-throughput data transmission \cite{ref55}. However, THz signals are highly susceptible to free-space path loss and molecular absorption, particularly under humid or adverse weather conditions \cite{ref56,ref57}.
   By integrating VLC and THz into a hybrid communication architecture, the system can dynamically switch between or combine both links. VLC provides high-speed, low-latency connectivity in favorable conditions, while THz serves as a fallback when optical channels are impaired. This complementary design improves data rates, minimizes outages, and improves reliability and resilience, key requirements for ITS applications in complex and high-mobility urban environments \cite{ref114,ref115}.}

   \textcolor{black}{ The full potential of the proposed VLC/THz hybrid system is realized only through a carefully engineered framework, where all components, from physical transceiver layout to adaptive switching mechanisms, are optimized to work together. In the following sections, we present a comprehensive design strategy, demonstrating how system-level optimizations converge to enable a robust, efficient, and deployment-ready I2V communication architecture for future ITS applications.}
   \vspace{-0.1cm}
\subsection{Literature Review}	
\textcolor{black}{Recent years have seen a surge of research  in the field of VLC, reflecting its growing importance and effectiveness (e.g., \cite{ref16
,ref42,ref46,ref67,ref69,ref71}). In particular, significant contributions have been made in vehicular VLC (VVLC), as highlighted in \cite{ref28,ref29,ref77,ref73,ref83,ref74,ref68,ref70,ref82,ref80,ref81}. Among these, \cite{ref28,ref74,ref70} primarily focus on V2V communication, with addressing channel modeling and optimal component placement.}
 A  review of existing works in V2V and I2V communications, focusing on the number of photodetectors (PDs) and transmitter types, is conducted in \cite{ref73}. Some I2V studies (e.g., \cite{ref75}) consider the PD position on the front hood of the vehicle, while others (e.g., \cite{ref76,ref77}) consider the top of the vehicle. Additionally, scattered efforts have been made for combined V2V and I2V communications \cite{ref78,ref79}.
	
In \cite{ref80}, non-line-of-sight (NLoS) transmission in VVLC communications is investigated. Also, in \cite{ref81} a multi-hop vehicular communication method using other vehicles is proposed for VVLC.  In \cite{ref29}, the authors examined the performance of I2V systems with access points shaped as streetlights. They also statistically analyzed the path loss and  bit error rate. In \cite{ref82}, it has first reminded the prohibition of using high-beam headlights in some cases and then investigated the system with a low-beam headlight transmitter. In \cite{ref83}, modeling of V2V-I2V systems has been performed. In \cite{ref77}, the effect of vehicle movement on information reception has been examined, and a dynamic soft handover algorithm has been proposed. In \cite{ref85}, intelligent reflecting surfaces are introduced as a promising technique for controlling wireless propagation environment to enhance communication performance.

In addition, some works on VLC/RF hybrid systems have 
been conducted.  In \cite{ref90}, a novel multiuser VLC/RF hybrid system is proposed, where users are paired: one receives data via VLC and relays it to the paired user through RF transmission. The performance of the outage probability of a hybrid VLC/RF IoT system was analyzed in \cite{ref92}, where VLC handles the downlink and RF is used for the uplink. In \cite{ref49}, a cognitive VLC/RF  hybrid system with decode and forward relaying was introduced for vehicular networks. This system employs a dual-link approach, initially selecting a direct link and switching to a hybrid link when signal quality degrades. 

Recent studies have explored the benefits of hybrid VLC/RF-based vehicle-to-everything (V2X) systems. In \cite{ref115}, the impact of interference and adverse weather conditions on link aggregated (LA)-aided  VLC/RF V2X systems is analyzed. Simulation results indicate  high reliability ($\sim$ 99.999$\%$ ) and low latency ($<$1 ms) within a 200 m range, even under challenging conditions. Additionally, \cite{ref114} examines VLC/THz communication links assisted by drones, demonstrating improvements in bit error rate and dropout probability, further emphasizing the potential of integrating these technologies.

A comprehensive literature review highlights a significant gap in research on VLC/THz hybrid systems and the holistic physical design of I2V communication systems that meet 6G requirements. While progress has been made in VLC and hybrid communications, most studies focus on isolated aspects, such as photodetector placement or specific transmission techniques, without integrating them into a unified I2V framework. Moreover, the design of a robust I2V system that ensures high reliability and low latency under adverse conditions, considering  mobility and environmental challenges, remains underexplored. To address this gap, we propose a foundational framework that builds upon our previous work~\cite{ref111} and introduces a novel, fully integrated approach potentially tailored to 6G-based ITS. Our key contribution is an optimized VLC/THz hybrid communication system designed to meet the stringent requirements of next-generation ITS networks. \textcolor{black}{To clearly highlight the novelty and scope of our proposed system relative to existing works, Table~\ref{jad0} provides a comparative summary of key features addressed across relevant studies.}
\vspace{-0.2cm}
\subsection{Contributions}
\textcolor{black}{In this paper, we complete the foundational structure of a proposed ITS for future urban environments by focusing on the design and analysis of the access network, physical layout of infrastructure components, and systematic parameter optimization. These efforts complement the previously proposed backhaul framework in \cite{ref111}. The key contributions are summarized as follows and are elaborated in the subsequent System Model section:}

    \begin{itemize}

    \item \textcolor{black}{\textbf{Design of a Hybrid VLC/THz Access Network:} We propose an access-layer architecture that integrates VLC with a THz antenna array system to enable high-speed, wide-coverage I2V communication.}

  \item \textcolor{black}{\textbf{End-to-End Access-Backhaul Integration:} We develop an access network architecture that seamlessly aligns with the backhaul layer, ensuring smooth, reliable I2V communication across the ITS. This integration guarantees robust, low-latency performance at high vehicular speeds.}

  \item  \textcolor{black}{ \textbf{Innovative Design of Physical Framework:} We propose a novel light‐pole architecture and mounting scheme that strategically positions and angles transmitters and receivers, maximizing roadway communication coverage while minimizing cross-link interference.}

   \item \textcolor{black}{ \textbf{Hybrid Switching-Combining Mechanism in the Access Layer:} We extend our previously proposed switching-combining (PSC) method to the access network, enabling transitions between VLC and THz links.}

  \item  \textcolor{black}{ \textbf{ Comprehensive System Parameter Optimization:}  A systematic optimization process is conducted across both physical and communication layers to maximize the combined performance of lighting and communication.}
\end{itemize}
The remainder of the paper is organized as follows: Section~II presents the system model. Section~III introduces the mathematical modeling of the VLC, THz, and hybrid systems and analyzes their performance. Section~IV discusses the simulation results. Finally, Section~V concludes the paper and outlines directions for future work.

		\begin{figure}[t]
		\centering
		\includegraphics[width=1\linewidth,height=0.2\textheight]{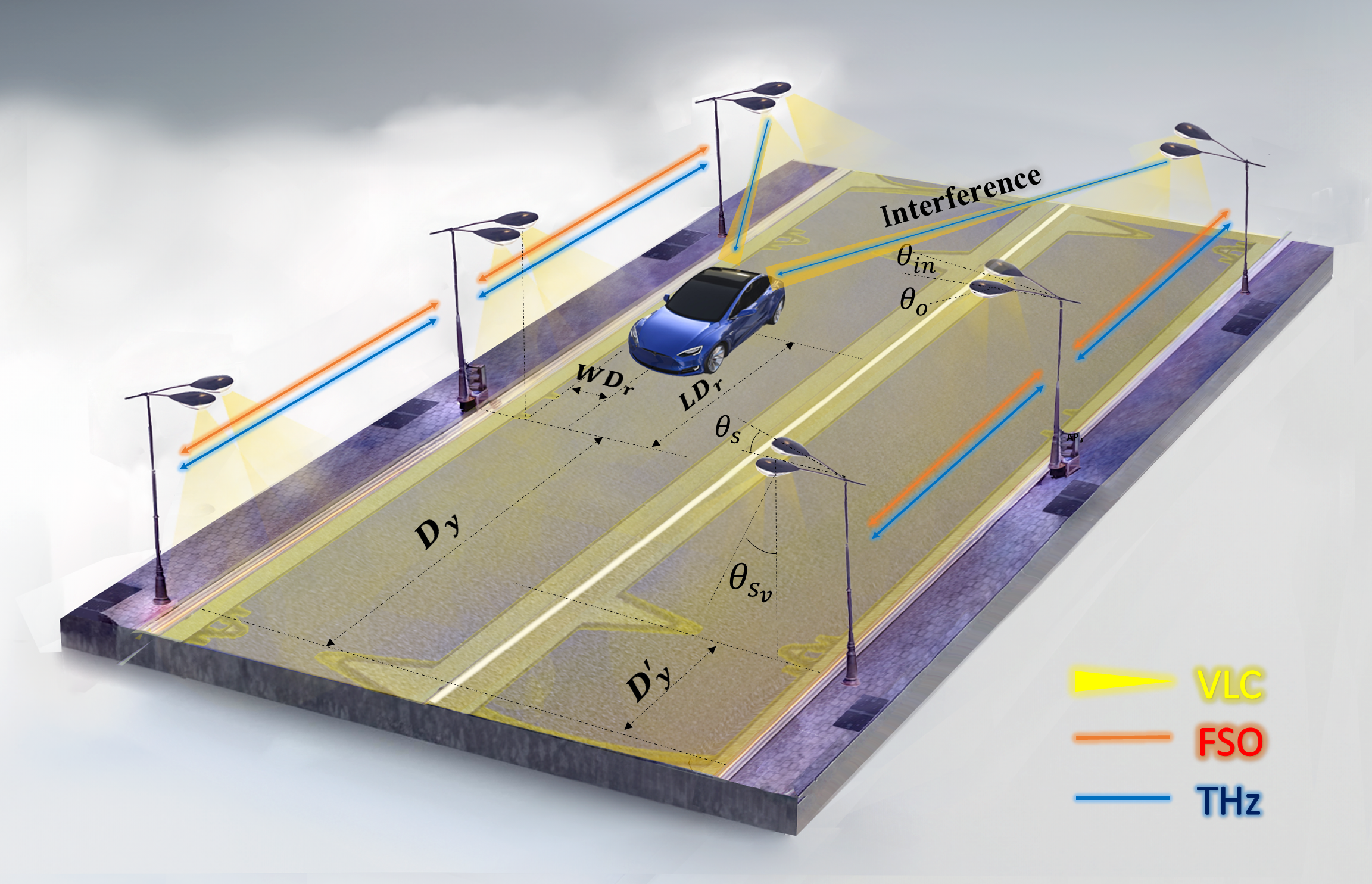}
		\caption{Overview of the proposed ITS architecture, featuring a hybrid FSO/THz system for the backhaul network and a hybrid VLC/THz system for the access network.}
		\label{fig3}
  \vspace{-0.4cm}
	\end{figure}

\section{System Model}
\textcolor{black}{ In this work, we develop an ITS architecture tailored for future smart cities. As shown in Fig.~\ref{fig3}, the proposed ITS framework follows a conventional two-layer communication architecture comprising a backhaul and an access layer. However, each layer incorporates novel technologies and innovative design strategies to enhance throughput, coverage, reliability, and low-latency performance. Since understanding the performance of the backhaul and its  integration with the access  is essential to evaluating the overall architecture, we first give a brief explanation of it.}
   \begin{figure}[t]
		\centering
             \vspace{-0.1cm}
		\includegraphics[width=0.8\linewidth,height=0.14\textheight]{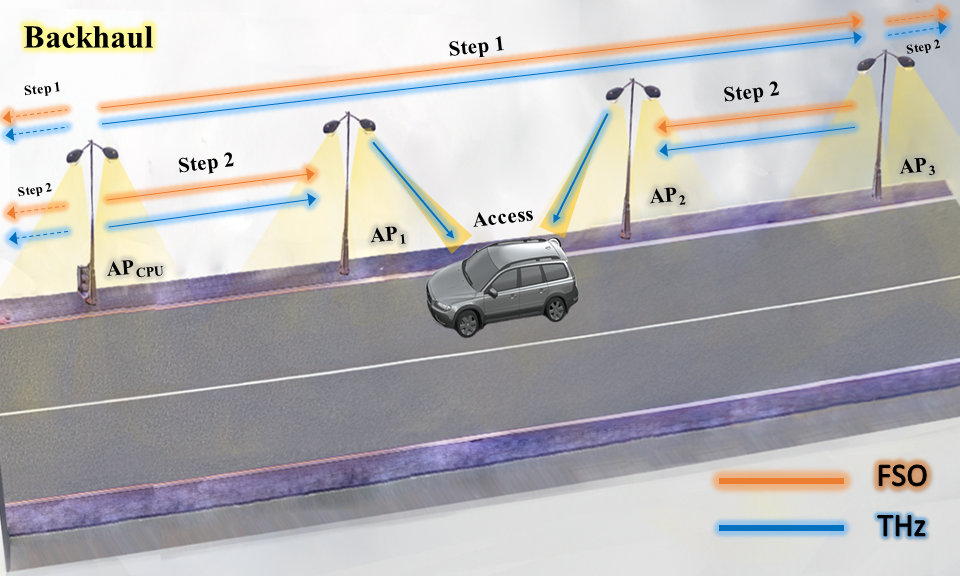}
          \vspace{-0.2cm}
		\caption{\textcolor{black}{A view of the backhaul layer of the system model \cite{ref111}.}}
                \vspace{-0.45cm}
		\label{fig0a}
	\end{figure}
 \vspace{-0.2cm}
 
\textcolor{black}{\subsection{Backhaul Layer}
 At the foundation, 
 the backhaul layer
 utilizes a hybrid free-space optical (FSO)/THz link to interconnect light poles.  This system employs a cell-free network of access points, represented by light poles to tackle the challenges posed by high-speed  vehicles, where latency, handover, and mobility are critical. By virtually eliminating cell boundaries, the cell-free network minimizes handover delays and ensures seamless connectivity. As depicted in Fig. \ref{fig0a}, an special light pole acts as an access point connected to the network core, equipped with its own control processing unit (CPU), and supports a chain of adjacent poles on both sides, extending to the next CPU-mounted pole. According to our prior findings \cite{ref111}, each CPU can reliably support up to 289 poles, covering  approximately 20 km of roadway. To further enhance data sharing speed  among the poles, we introduced a dedicated method  that reduces delay by nearly one-third, thereby enabling rapid and low-latency data dissemination across the entire network.}

 \vspace{-0.03cm}
 \vspace{-0.55cm}
\textcolor{black}{\subsection{Access Layer}
Located along the backhaul layer is the access network designed to establish a robust and reliable connection between the ITS infrastructure and its users, specifically autonomous vehicles. Each streetlight pole serves as a multifunctional node, simultaneously delivering efficient road illumination and high-performance communication coverage. Through cooperative communication among adjacent poles, the system effectively reduces coverage gaps, ensuring near-continuous connectivity across the roadway.
To demonstrate the system's capabilities, envision a forward-looking scenario in which each vehicle entering the smart roadway is automatically assigned a unique IP address upon detection. This allows the network core to maintain real-time awareness of each vehicle’s location, speed, and trajectory. Leveraging this centralized information, the system constructs and continuously updates a dynamic traffic map, which is distributed across all network nodes. This enables a broad range of intelligent services such as real-time traffic flow, collision avoidance, and adaptive route guidance.}


\textcolor{black}{Beyond the intelligent communication architecture, this work introduces a series of physical design innovations aimed at enhancing installation efficiency, interference mitigation, and energy performance. As illustrated in Fig. \ref{fig3}, each smart light pole is engineered to ensure optimal placement and orientation of transmitters and receivers. These design choices improve both illumination uniformity and communication coverage, while also contributing to reduced power consumption and lower infrastructure costs.
A critical component of the physical design is interference management. To address signal disruption common in two-way road configurations, transmitters and receivers are strategically aligned to minimize cross-lane interference. This layout ensures stable, high-quality links even in dense urban environments. }

\textcolor{black}{In addition to architectural innovations, we perform a systematic optimization of key control parameters to enhance overall system performance. This includes precise adjustments to the longitudinal placement and angular orientation of transceivers, as well as communication settings for both VLC and THz subsystems. By conducting a comprehensive grid search over thousands of parameter combinations, we identify configurations that strike an optimal balance between illumination uniformity, link quality, and hybrid communication efficiency. Results presented in Section IV demonstrate significant improvements in coverage, reliability, and energy savings, highlighting the benefit of coordinated physical and communication-layer optimization for next-generation ITS.}

\textcolor{black}{ Another critical component for ensuring consistent and resilient performance  under varying environmental and operational conditions in the proposed ITS framework is the switching-combining method, designed to intelligently manage the dynamic interplay between  communication systems across both the backhaul and access network segments. In the backhaul, the system leverages FSO/THz point-to-point links, while the access layer integrates VLC with a distinct THz system utilizing array antennas for broad, area-wide coverage. As demonstrated by our results, this mechanism plays a crucial role in optimizing the performance of hybrid systems across both   segments. A detailed explanation of its design and functionality is provided in the corresponding subsection.}

\textcolor{black}{
 By combining cell-free network design, high-speed technologies (VLC, THz, and FSO), innovative methods in switching and data transfer mechanisms, and an optimized physical design, this framework establishes a scalable, resilient, and future-proof communication backbone. These innovations collectively position the system as a viable solution for the demanding communication needs of smart cities, particularly in support of autonomous and connected vehicle ecosystems. Fig. \ref{fig3} presents a comprehensive overview of the proposed ITS framework, illustrating how existing street lighting infrastructure is leveraged to enable seamless I2V communication. In this figure, the access network is highlighted, including street geometry such as a lane width of  $WD_l$ and light pole height of $H_t$. The following subsections provide a detailed description of the transmitter and receiver designs, tailored specifically to the performance and coverage needs of our application.}
 \vspace{-0.1cm}

	\subsection{Transmitter}
	As mentioned,  we have custom-designed a structure for streetlights tailored to our application, assuming that each of the light poles carries two lights at different angles. Thus, the top view of the proposed structure resembles Fig. \ref{fig4}. The advantage of this structure is that it increases the number of parameters available for controlling the coverage of both lighting and communication and allows us to increase the longitudinal distance between streetlights. The lateral and longitudinal distances of each light from the center of the light pole are obtained as:
    \vspace{-0.02cm}
    	\begin{figure}[!h]
		\centering
		\includegraphics[width=0.7\linewidth,height=0.14\textheight]{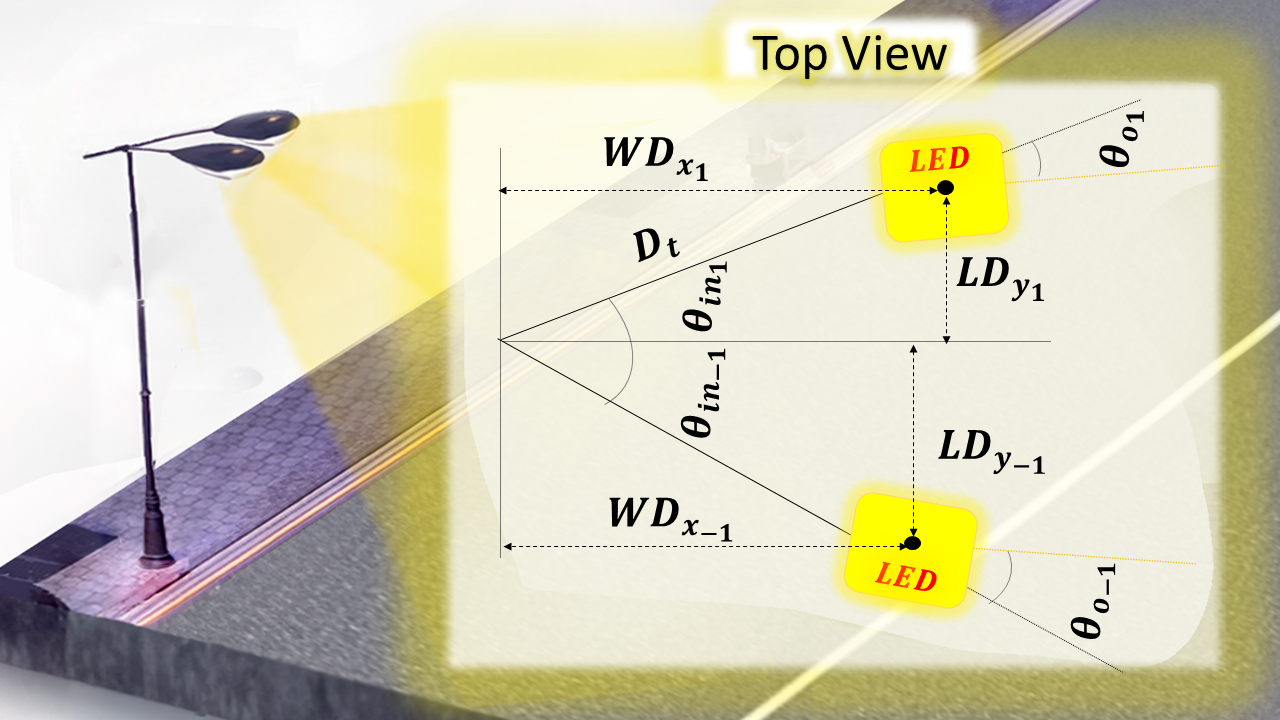}
		\caption{The proposed structure of street lights.}
		\vspace{-0.4cm}
		\label{fig4}
	\end{figure}	
	\begin{equation}
			\begin{gathered}
	{	\left\{ \begin{array}{ll}
			 W D_{x_k}=\frac{d_S}{2} \cos \left(\theta_{i n_k}-\theta_{o_k}\right)+D_{t} \cos \theta_{i n_k}, & k=-1,1\\
		L D_{y_k}=\frac{d_S}{2} \sin \left(\theta_{i n_k}-\theta_{o_k}\right)+D_{t} \sin \theta_{i n_k}, & k=-1,1,
				\end{array} \right.} \\
		\end{gathered}
		\label{eq1}
	\end{equation}
	where $ d_s $ is the side length of the transmitter's square surface. $ \theta_{in} $ shows the angle between the line parallel to the transverse axis of the street, which passes through the center of the light pole, and each of the two rods holding the lights (with the length $ D_{t} $). $ \theta_{o}  $ is the angle between the direction of the light  and its holding rod. $ WD_{x_k} $ and $ LD_{y_k} $ represent the lateral and longitudinal distances, respectively, from the light's center to the pole's center. When viewed from the center of the pole, $ k=-1 $ refers to the right-side transmitter, and $ k=1 $ refers to the left-side transmitter. Note that the physical structure of the location of THz transmitters also follows this structure.
	\begin{figure}[!h]
		\centering
             \vspace{-0.1cm}
		\includegraphics[width=0.7\linewidth,height=0.13\textheight]{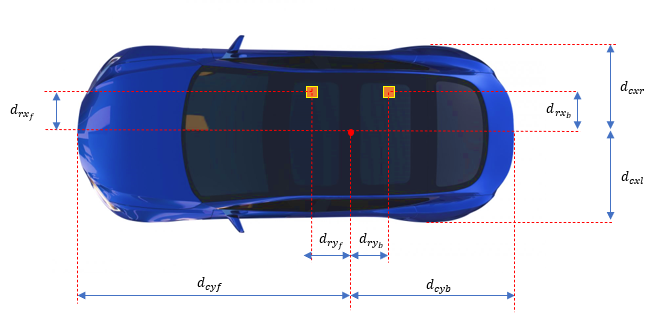}
          \vspace{-0.2cm}
		\caption{Structure and location of detectors on the car roof.}
                \vspace{-0.3cm}
		\label{fig5}
	\end{figure}
	\subsection{Receiver}	
   For lighting coverage, the receivers are modeled as points on the street surface, each represented by a square with a defined side length. In the vehicle communication scenario, however, two receivers are mounted on the vehicle's roof for both the VLC and THz systems, as depicted in Fig.~\ref{fig5}. To maximize received power, the receivers are positioned at the maximum possible distance from each other, while both are placed along the right edge of the vehicle to minimize interference. This configuration significantly enhances the reception of the highest useful power while reducing interference. The specific parameter values shown in the figure are summarized at the bottom of Table~\ref{jad2}.
   
    \begin{figure}[!h]
    \vspace{-0.5cm}
    \subfloat[\label{fig5_0a}]{%
        \includegraphics[width=0.5\linewidth,height=0.14\textheight]{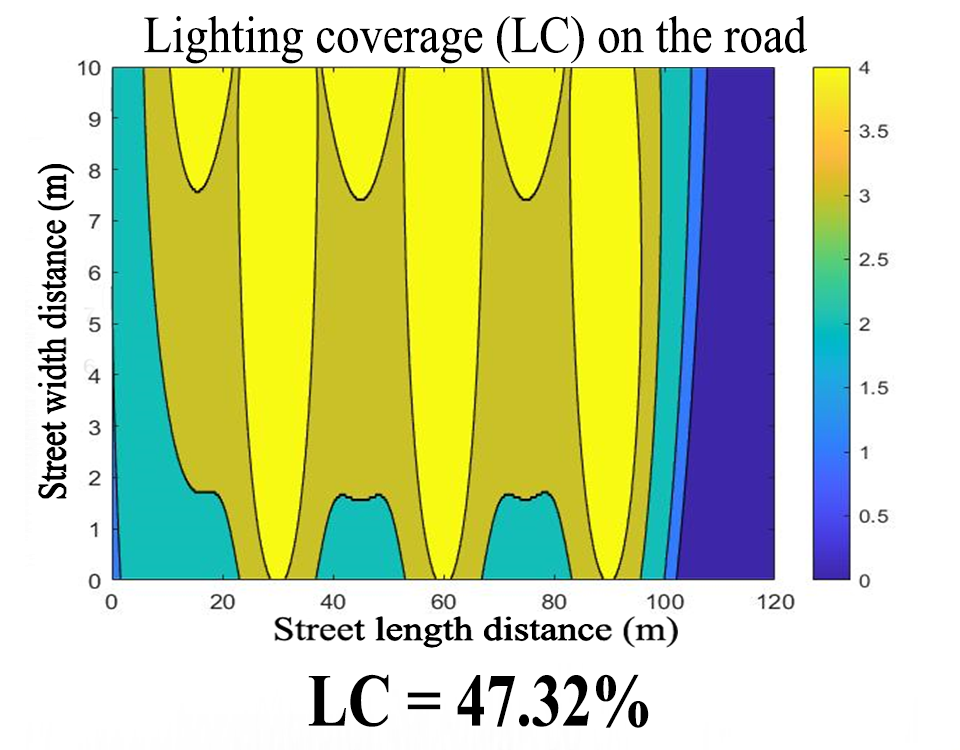}%
    }
    \subfloat[\label{fig5_0b}]{%
        \includegraphics[width=0.5\linewidth,height=0.14\textheight]{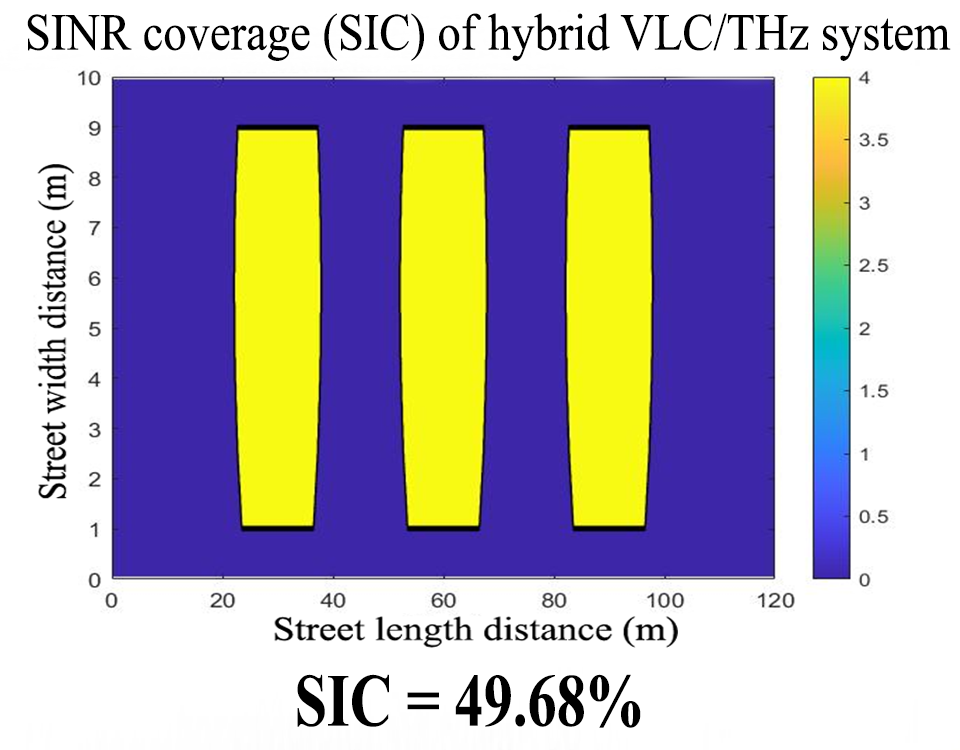}%
    }
    \caption{Coverage on the street surface using the general standard configuration: (a) lighting coverage (LC), (b) SINR coverage (SIC) of the hybrid system.}
    \vspace{-0.4cm}
    \label{fig:fig5_0}
\end{figure}

\textcolor{black}{To demonstrate the practical significance and experimental feasibility of our dedicated transmitter and receiver design, Fig.~\ref{fig:fig5_0} is presented. In Fig.~\ref{fig5_0a}, the lighting coverage is illustrated, while Fig.~\ref{fig5_0b} shows the communication coverage of the hybrid system under a baseline configuration aligned with real-world urban planning standards. In this reference scenario, each light pole is equipped with a single lighting unit and spaced 30 meters apart, and the receiver is mounted at the center of the vehicle roof, consistent with common experimental setups in the literature (e.g., \cite{ref29,ref77,ref83}). Moreover, only the basic angular parameters $\theta_S$ and $\theta_R$ are applied to the transmitters and receivers, respectively, using power levels listed in Table~\ref{jad2}. As shown, even after optimizing available parameters within this conventional configuration, the maximum achievable coverage is limited to 47\% for lighting and 49\% for hybrid communication.
These findings highlight real deployment limitations such as misalignment sensitivity and limited beam coverage. To address such challenges, our proposed structure includes experimentally informed design enhancements, including custom beam orientations and receiver alignments, which are discussed in detail in this section. With these modifications, we not only extend the pole spacing to 50 meters, substantially lowering infrastructure cost, but also achieve 97\% lighting coverage and 99.9\% hybrid communication coverage, as illustrated in the results section. Importantly, attaining similar performance using the baseline design would require nearly three times more power, further emphasizing the efficiency and practical viability of our design for real-world deployment scenarios.}
     \begin{figure}[!h]
		\centering
             \vspace{-0.2cm}
		\includegraphics[width=0.45\linewidth,height=0.14\textheight]{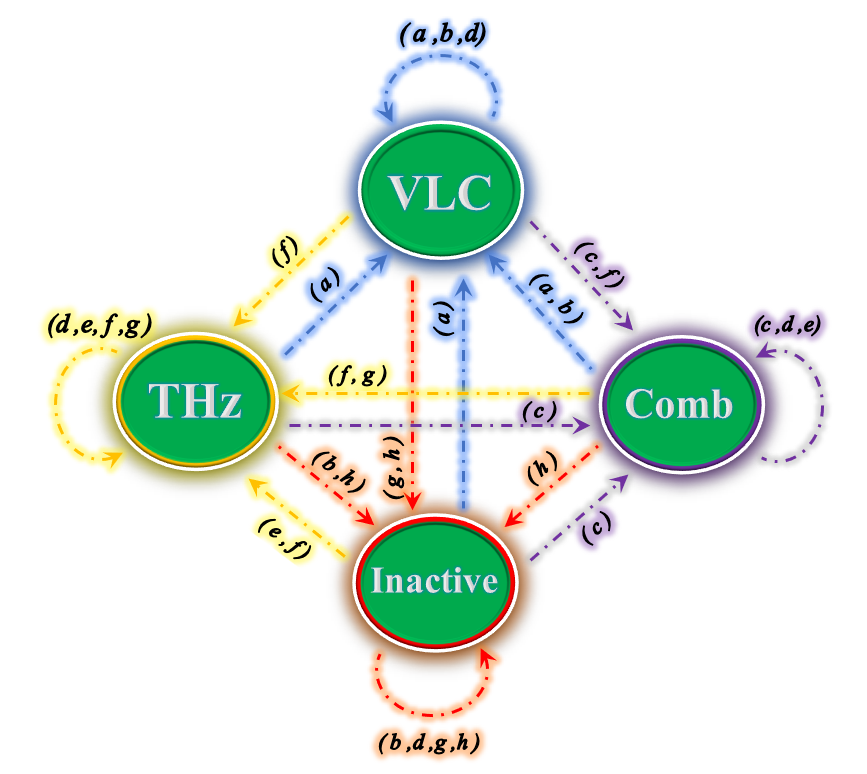}
          \vspace{-0.3cm}
		\caption{\textcolor{black}{Block diagram of the switching-combining mechanism in the access}.}
                \vspace{-0.5cm}
		\label{fig0b}
	\end{figure}
\subsection{\textcolor{black}{Switching-Combining Mechanism}}

\textcolor{black}{Reliable operation of a hybrid communication system critically depends on an effective link adaptation or switching strategy. The simplest method, general switching (GS), uses a fixed threshold to trigger a switch between technologies  when signal quality drops below the limit. However, such  approaches often lead to frequent switching, which can degrade performance, increase latency, and reduce hardware longevity. To mitigate these drawbacks, soft switching (SoftSw) schemes have been introduced \cite{ref62}, enabling smoother transitions by gradually adjusting link usage. Meanwhile, combining techniques (e.g., \cite{ref56}) aim to improve reliability by simultaneously activating both communication links. Though effective in harsh conditions, this strategy consumes more power and may be inefficient under normal operation.}

\textcolor{black}{To strike a balance between robustness and efficiency, we design a proposed switching-combining (PSC) mechanism. This mechanism adaptively integrates both switching and combining based on real-time system and environmental conditions. As visualized in Fig.~\ref{fig0b} and detailed in \cite{ref111}, it evaluates a comprehensive set of criteria (conditions a–h) to decide whether to: (i) maintain the current mode, (ii) switch between technologies, or (iii) activate both simultaneously. To further enhance performance and reduce unnecessary switches, the mechanism incorporates two thresholds, defined upper and lower bounds that introduce stability by preventing oscillations between modes. As shown in the results section, this approach offers superior performance in comparison with traditional GS, SoftSw, and  combining strategies.}

\textcolor{black}{ From a latency perspective, the switching process involves three stages: (1) receiving the SINR/SNR signal from the vehicle (propagation delay), (2) decision-making through threshold comparison, and (3) execution of the switching action. Given the short communication distance ($\le$ 100 m), the propagation delay is limited to approximately 0.5 µs, the  decision-making, if implemented using FPGA or MCU, adds less than 1 µs, and according to \cite{ref118}, the  switching time  is below 5 µs. As a result, the total switching latency remains under 7 µs, ensuring compatibility with delay-sensitive ITS applications. More importantly, since the total switching delay is primarily governed by the third stage, the reduced number of switches achieved by our PSC strategy plays a critical role in minimizing the overall system latency.}

	\section{Performance Analysis}
		
	\subsection{ VLC Received Power}
	
We assume that the coordinate origin is located at the starting point of the asphalt, with the longitudinal axis of the street represented by \( y \) and the transverse axis by \( x \). As shown in Fig. \ref{fig3}, we consider \( K=3 \) streetlights spaced at a distance of \( D_y \) from one another, resulting in a street that extends from the origin to a length of \( LD_l = (K+1) \times D_y \). The longitudinal and lateral positions where the center of a vehicle can be located are stored in the vectors \( WD_r \) and \( LD_r \), respectively. There are no streetlights positioned at \( y=0 \) or \( y=(K+1) \times D_y \), and the first streetlight is placed at \( y=D_y \).

Each transmitter and receiver is characterized by four angles: the first two angles, \( \theta_{o} \) and \( \theta_{in} \), were introduced in Section II. The third angle represents the inclination of the transmitter or receiver surface relative to the horizontal surface, denoted as \( \theta_{S} \) for transmitters and \( \theta_{R} \) for receivers. These angles are assigned to pairs of transmitters mounted on light poles and pairs of receivers mounted on vehicles, with their values stored in corresponding two-element vectors. The fourth angle corresponds to the adjustment made for the receiver located further back relative to the car’s previous normal vector, tilting it towards the rear light. Similarly, the forward-facing receiver adjusts towards the streetlight ahead. These angles are recorded in the \( \theta_{R_{v}} \) vector. The same principle applies to the transmitters on the light poles: the right-side light tilts more to the right by the angle stored in the \( \theta_{S_{v}} \) vector, while the left-side light tilts accordingly. These adjustments are also depicted in Fig. \ref{fig3}.
    	
	\begin{figure}[!t]
		\centering
		\includegraphics[width=1\linewidth,height=0.17\textheight]{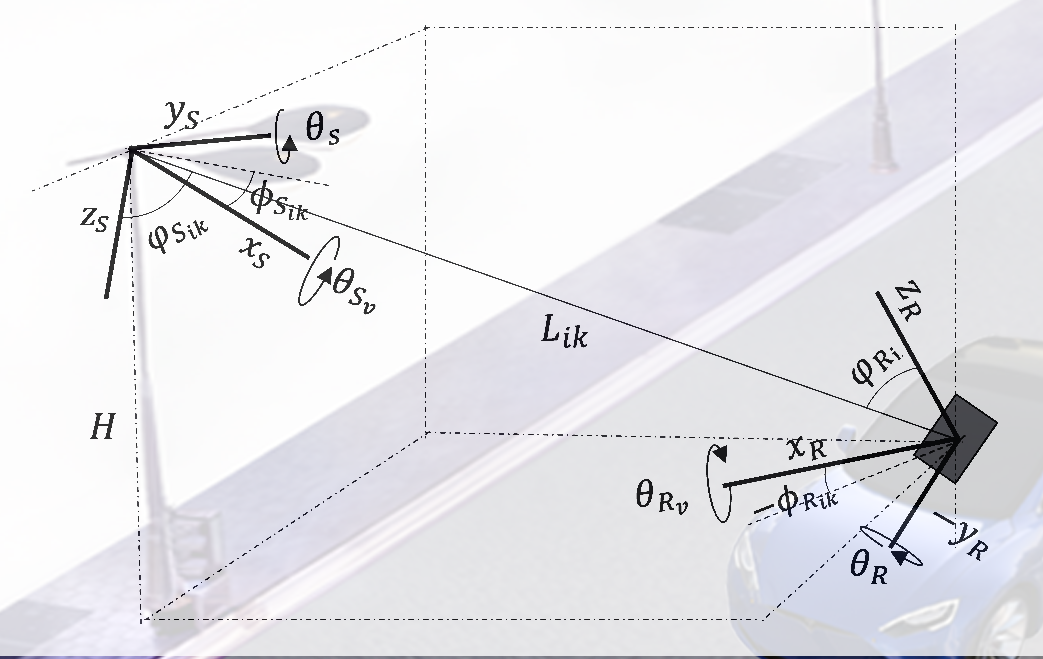}
  \vspace{-0.6cm}
		\caption{Schematic representation of the physical structure and  communication modeling  between a transmitter and a receiver, illustrating all of the angles.}
        \vspace{-0.5cm}
		\label{fig6}
	\end{figure}	
	For modeling this scenario, we use the geometric method approved in the communication systems literature. This method has been used in \cite{ref109}. Fig. \ref{fig6} shows a schematic representation of the mentioned method, where for the front receiver, the car's roof, and the transmitter on the left side of the streetlight are considered. 	
	This method states that if we consider a point $ (x_0, y_0, z_0) $ in a coordinate system and apply rotations on each of the axes $ (x, y, z) $, the transformations $ R_x, R_y, R_z $ are applied to this point, and the resulting points $ (x_0^{\prime}, y_0^{\prime}, z_0^{\prime}) $ are obtained according to the new coordinate axes in  \eqref{eq2} as follows:
    \begin{equation}
		{\left(x_0^{\prime}, y_0^{\prime}, z_0^{\prime}\right)^T=R_z\left(\theta_z\right) R_y\left(\theta_y\right) R_x\left(\theta_x\right)\left(x_0, y_0, z_0\right)^T },
		\label{eq2}
	\end{equation}
    where
	$$
	\begin{aligned}
		& R_x\left(\theta_x\right)=\left(\begin{array}{llr}
			1 & 0 & 0 \\
			0 & +\cos \theta_x & -\sin \theta_x \\
			0 & +\sin \theta_x & +\cos \theta_x
		\end{array}\right),\\
		& R_y\left(\theta_y\right)=\left(\begin{array}{llr}
			+\cos \theta_y & 0 & +\sin \theta_y \\
			0 & 1 & 0 \\
			-\sin \theta_y & 0 & +\cos \theta_y
		\end{array}\right),\\
		& R_z\left(\theta_z\right)=\left(\begin{array}{lll}
			+\cos \theta_z & -\sin \theta_z & 0 \\
			+\sin \theta_z & +\cos \theta_z & 0 \\
			0 & 0 & 1
		\end{array}\right). \\
	\end{aligned}
	$$
    
  \textcolor{black}{  As illustrated in Fig.~\ref{fig6}, the rotation angles for the transmitter coordinate axes are defined as: 
\(\theta_{z_S} = 0\), \(\theta_{y_S} = -\theta_S\), and \(\theta_{x_S} = k \times \theta_{S_v}\). 
Similarly, for the receiver, the angles are: 
\(\theta_{z_R} = 0\), \(\theta_{y_R} = -\theta_R\), and \(\theta_{x_R} = i \times \theta_{R_v}\). 
Assuming the coordinate origin is located at the transmitter \((j = -1,\, M = S)\) or at the receiver \((j = 1,\, M = R)\), the initial coordinates of the corresponding receiver/transmitter are:}
	\begin{equation}
			\begin{gathered}
	{	\left\{ \begin{array}{l}
			 x_{0 M_{i k}}=W D_r-i \times d_{r x_i}-W D_{x_k}, \\
			 y_{0 M_{i k}}=j \times \left(L D_r+i \times d_{r y_i}-k \times L D_{y_k}\right), \\
			 z_{0 M_{i k}}=H=H_t - H_r .  \\
			\end{array} \right.} \\
		\end{gathered}
		\label{eq3}
	\end{equation}
    
\textcolor{black}{   The transformed coordinates after applying the rotations are computed as:}
	\begin{equation}
 \begin{gathered}
 {	\left\{ \begin{array}{l}
			 {x}_{ {0}  {M}_{ {i}  {k}}}^{\prime}\!= {x}_{ {0}  {M}_{ {i k}}}\!\!\left(\cos \theta_{\!M} \!\cos \theta_{{\!M}_{v }}\!\right)+ {y}_{ {0}  {M}_{ {i}  {k}}}\!\!\left(i  \sin \theta_{\!M} \sin \theta_{{\!M}_{v }}\!\right) \\
			 \quad\quad\;\quad + {z}_{ {0}  {M}_{ {i k}}}\!\left(\sin \theta_M \cos \theta_{M_{v }}\right), \\
			 {y}_{ {0} {M}_{ {i}  {k}}}^{\prime} = {y}_{ {0}  {M}_{ {i k}}}\left(\cos \theta_{M_{v }}\right)+ {z}_{ {0}  {M}_{ {i k}}}\left(-i  \sin \theta_{M_{v }}\right), \\
			  {z}_{ {0} {M}_{ {i k}}}^{\prime} = {x}_{ {0}  {M}_{ {i k}}}\left(-\sin \theta_M\right)+ {y}_{ {0}  {M}_{ {i k}}}\left(i  \cos \theta_M \sin \theta_{M_{v }}\right)\\
			 \quad\quad\;\quad + {z}_{ {0}  {M}_{ {i k}}}\left(\cos \theta_M \cos \theta_{M_{v }}\right). \\	
  		\end{array} \right.} \\
		\end{gathered}
		\label{eq4}
	\end{equation}
\textcolor{black}{here, \(\theta_M\) and \(\theta_{M_v}\) are the rotation angles shown in Fig.~\ref{fig6}, initially defined at the beginning of this section and further clarified in Table~\ref{jad2}. In that table, the subscript ‘V’ refers to VLC system configurations, while ‘T’ refers to THz system settings. The parameters \(W D_r\) and \(L D_r\) are defined earlier in this section. The index \(k\) is introduced in the transmitter subsection, and \(i \in \{-1, 1\}\) represents the two receivers placed on the vehicle’s roof. Thus, we achieve the transmitting and receiving angle and the   link length as:}
	\begin{equation}
 \begin{array}{cc}
		{ L_{i k}=\sqrt{x_{0R_{i k}}^{\prime\;2}+y_{0R_{i k}}^{\prime\;2}+z_{0R_{i k}}^{\prime\;2}}} \; ,\quad i=-1,1 , \quad k=-1,1 \\
  { \varphi_{R_{i k}}=\cos ^{-1} \frac{Z_{0R_{i k}}^{\prime}}{L_{i k}} } ,\qquad\;\;
  { \varphi_{S_{i k}}=\cos ^{-1} \frac{Z_{0S_{i k}}^{\prime}}{L_{i k}}  }.
  \end{array}
  	\label{eq5}
	\end{equation}
    
       Following the approach used in the references (such as in \cite{
       ref71,ref40,ref102,ref103}), and assuming Lambertian transmitters, we can obtain the received power of the vehicle at each point on the street surface from a pair of transmitters installed on a light pole  as:     
       \begin{equation}	 
       {P_c=\sum_{k, i=-1,1} P_{i k}} 	,
       \label{eq8}
	\end{equation}
 where
	\begin{equation*}
		\begin{gathered}
			{	 P_{i k}\!=\!\left\{\begin{array}{ll}
					\!\!\!\!P_t\frac{  A_R T_s\; g \;(m+1)h_{a a} \cos ^m \!\!\varphi_{S_{i k}} \!\!\cos \varphi_{R_{i k}}}{2 \pi L_{i k}^2} \;, & \begin{array}{l} \!\!0 \leq\!\!  {\varphi}_{R_{i k}} \!\!\leq  {\varphi}_{R_c} \\ \!\!0 \leq  \!{\varphi}_{S_{i k}}\!\! \leq  {\varphi}_{S_c} \end{array}\\
				0 \;, &  o.w ,
				\end{array}\right.} \\
		\end{gathered}
	\end{equation*}	
 where \( m \) is the Lambert coefficient, \( P_t \) denotes the transmitter output power, and \( A_S \) and \( A_R \) represent the areas of the transmitter and receiver, respectively. The angle \( \varphi_S \) denotes the angle between the normal vector of the transmitter plane and the line connecting the transmitter and receiver, referred to as the transmitting angle, while \( \varphi_R \) denotes the same for the receiver plane, referred to as the receiving angle. The link length is represented by \( L \), \( T_s \) represents the filter pass response, and \( g \) is referred to as the concentrator gain. We assume both \( T_s \) and \( g \) to be unity by default. 

Furthermore, \( \varphi_{R_c} \) and \( \varphi_{S_c} \) represent the fields of view. The term \( P_{ik} \) indicates the received power at each of the two receivers mounted on the roof of the car from each light mounted on a light pole. Consequently, \( P_c \) represents the total power received by the car from a light pole. Finally, in Equation \eqref{eq8}, \( h_{aa} \) represents the atmospheric attenuation effect, which can be modeled in the environment using the Beer-Lambert law as:

	\vspace{-0.4cm}
	\begin{equation}
		{	h_{a a}=\exp \left(-\alpha_a L\right) },
		\label{eq9}
	\end{equation}
	where  $\alpha_a$ represents the attenuation coefficient, which can be obtained as:
	\begin{equation}
		\alpha_a=\frac{3.912}{Vi}\left(\frac{\lambda_{Vi}}{550}\right)^{\delta\left(Vi\right)},
		\label{eq10}
	\end{equation}
	\[
	\delta\left(Vi\right)=\left\{\begin{array}{lr}
		1.6  \;, & \text{for}~ Vi>50 \mathrm{~km} \\
		1.3  \;,&  \text{for}~  6 \mathrm{~km}<Vi<50 \mathrm{~km} \\
		0.16 Vi+0.34  \;,&  \text{for}~  1 \mathrm{~km}<Vi<6 \mathrm{~km} \\
		Vi-0.5  \;,& \text{for}~  0.5 \mathrm{~km}<Vi<1 \mathrm{~km} \\
		0  \;,&  \text{for}~  Vi<0.5 \mathrm{~km},
	\end{array}\right.
	\]
	where $\lambda_V$ is the VLC wavelength and visibility ($Vi$) is a parameter that is determined based on weather conditions.
 
	\subsection{THz Received Power}
	In the intended communication system, we will use array antennas as transmitters and receivers. The corresponding array factor normalized equation is as \cite{ref112}:
	\begin{equation}
		A F_n(\varphi, \phi)=\left\{\frac{1}{N_x} \frac{\sin \left(\frac{N_x}{2} \psi_x\right)}{\sin   \left(\frac{\psi_x}{2}\right)}\right\}\left\{\frac{1}{N_y} \frac{\sin \left(\frac{N_y}{2} \psi_y\right)}{\sin \left(\frac{\psi_y}{2}\right)}\right\} ,
		\label{eq11}
	\end{equation}
	\vspace{-0.2cm}
	\begin{equation*}
		\begin{aligned}
			\psi_x=k d_x \sin \varphi \cos \phi+\beta_x, &\qquad\qquad\!\!\beta_x=-k d_x \sin \varphi_0 \cos \phi_0,\\
			\psi_y=k d_y \sin \varphi \sin \phi+\beta_y, &\qquad\qquad\!\! \beta_y=-k d_y \sin \varphi_0 \sin \phi_0, \\
		\end{aligned}
	\end{equation*}
 where \( N_x \) and \( N_y \) represent the number of antenna elements along the \( x \)-axis and \( y \)-axis, respectively, and are organized into two-element vectors for the transmitter and receiver, denoted as \( N_S \) and \( N_R \). The angle \( \varphi \) specifies the transmitting or receiving angle, similar to the VLC system, while \( \phi \) represents the imaging angle of the line connecting the transmitter and receiver on the respective planes. Additionally, \( \beta_x \) and \( \beta_y \) denote the phase shift parameters, where \( \phi_0 \) and \( \varphi_0 \) indicate the main lobe angles. 
	\begin{figure}[H]
       \vspace{-0.2cm}
		\centering
		\includegraphics[width=0.8\linewidth,height=0.15\textheight]{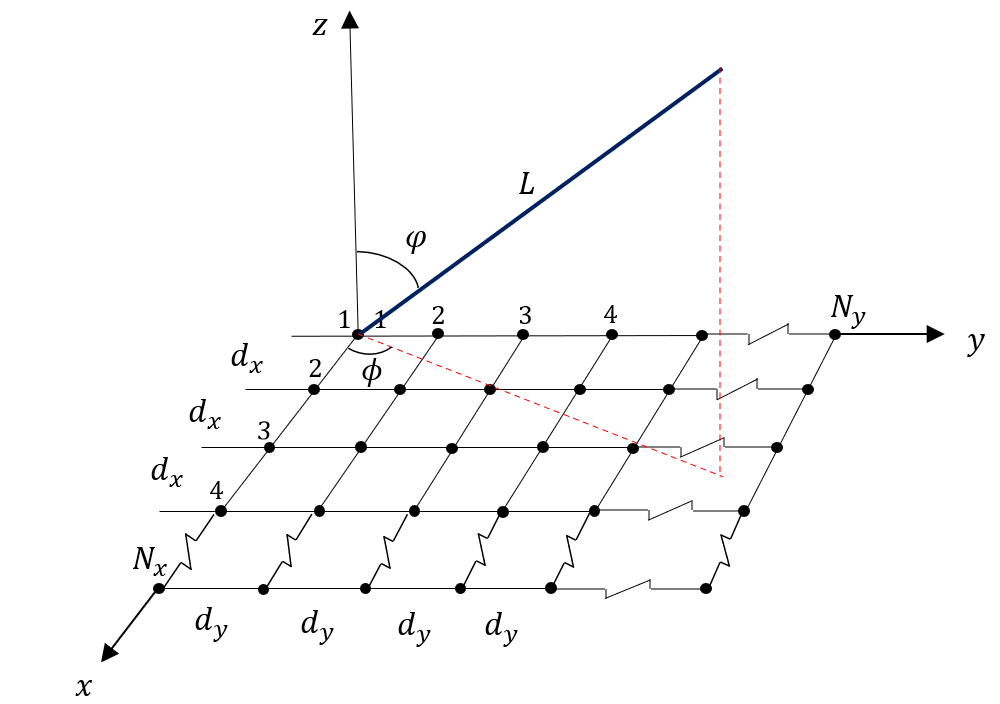}
           \vspace{-0.2cm}
		\caption{Conceptual representation of array antennas in THz system.}
        \vspace{-0.2cm}
		\label{fig7}
	\end{figure}
    The element spacing along the \( x \)- and \( y \)-axes is given by \( d_x = d_y = \frac{\lambda_T}{2} \), where \( \lambda_T \) is the THz wavelength, and \( k = \frac{2\pi}{\lambda_T} \) represents the wavenumber. Fig. \ref{fig7} illustrates the structure of these antennas.
	We previously derived the equations for \( \varphi \) and \( L \) in \eqref{eq5}, and the same principles apply here for the THz system. Therefore, the only additional calculation required is for \( \phi \). Using a similar approach, the values for the transmitting and  receiving angles \( \phi \) are obtained as follows:
	\begin{equation}
	\phi_{S_{i k}}=\tan ^{-1} \frac{y_{ {0} S_{i k}}^{\prime}}{x_{ {0} S_{i k}}^{\prime}} , \qquad
		\phi_{R_{i k}}=\tan ^{-1} \frac{y_{ {0} {R}_{i k}}^{\prime} }{ {x}_{ {0} R_{{i k}}^{\prime}}}. \\
		\label{eq13}
	\end{equation}
	
	Using \eqref{eq13}, the received power is modeled as: 
	\begin{equation}	
 {P_{th}=\sum_{k, i=-1,1} P_{th_{i k}}},
 	\label{eq14}
 \end{equation}
 where
	\begin{equation*}
		\begin{gathered}
			{P_{th_{i k}}=\left\{\begin{array}{cc}
					P_t h_{pl}^2 G_{S_{i k}} G_{R_{i k}} \;,& \begin{array}{c} 0 \leq \varphi_{R_{i k}}, \leq \varphi_{R_c} \\ 0 \leq \varphi_{S_{i k}}, \leq \varphi_{S_c} \end{array}\\
					0 \;,&  o.w, 
				\end{array}\right.} \\
		\end{gathered}
	\end{equation*}
 and
	\begin{equation*}
		\begin{gathered}
	{	\left\{ \begin{array}{l}
			G_{S_{i k}}\left(\varphi_{S_{i k}}, \phi_{S_{i k}}\right) \simeq \pi (N_{S y} N_{S x} A F_n\left(\varphi_{S_{i k}}, \phi_{S_{i k}}\right))^2\;,\\
			G_{R_{i k}}\left(\varphi_{R_{i k}}, \phi_{R_{i k}}\right) \simeq \pi (N_{R y} N_{R x} A F_n^2\left(\varphi_{R_{i k}}, \phi_{R_{i k}}\right))^2 \; ,  \\
   \end{array} \right.} \\
		\end{gathered}
	\end{equation*}
	For THz channel modeling, we adopt the approach outlined in \cite{ref55,ref56,ref58,ref60,ref62,ref71,ref106,ref107}. Let \( h_T = h_{pl} h_{ssf} h_{mf} \) represent the THz channel coefficient, where \( h_{ssf} \) accounts for small-scale fading, \( h_{mf} \) denotes the misalignment error, and \( h_{pl} \) represents the path loss. Since our focus is on achieving coverage across the entire street surface rather than point-to-point communication, the effect of misalignment error can be neglected. Additionally, given that the communication link length is only several tens of meters, the impact of small-scale fading can also be reasonably disregarded \cite{ref113}.

The path loss in the THz channel is expressed as the product of two factors, \( h_{pl}(t) = h_{pl_{free}}(t) h_{pl_{mol}}(t) \), where \( h_{pl_{free}}(t) \) is the free-space path loss and \( h_{pl_{mol}}(t) \) represents the molecular absorption loss, calculated using the Beer–Lambert law. By multiplying these two components, the total path loss coefficient for the THz channel is obtained as:
\vspace{-0.01cm}
	\begin{equation}
		{	h_{p l}(t)=\frac{c }{4 \pi f_t L} \exp \!\left(\!-\frac{1}{2} L\left(\sum_i M_i\left(f_t, \rho\right)+N\left(f_t, \rho\!\right)\right)\!\right) }.
		\label{eq17}
	\end{equation}
	where $ f_t $ is the THz frequency, and $ c $ is the speed of light.  $\rho=\frac{r_h P_{press }(T, p)}{100 p}$ is the volumetric mixing ratio of water vapor, where $ r_h $ is the relative humidity, $ T  $ is the ambient temperature, $ p  $ is the pressure,  and 
	$ P_{press}(T, p)=6.1121\left(1.0007+3.46 \times 10^{-6} p\right) \exp \left(\frac{17.502 T}{240.97+T}\right) $, 
	represents the saturation vapor pressure. Additionally, $ M_i (f_t ,\rho)  $ represents the polynomial expression for the six main absorption lines with central frequencies of 119, 183, 325, 380, 439, and 448 gigahertz. The equations related to $ M_i(f_t,\rho) $ and  $ N(f_t,\rho) $ are given as: 
$$
 { N\left(f_t, \rho\right)=\frac{\rho}{0.0157}\left(2 \times 10^{-4}+0.915 \times 10^{-112} f_t^{9.42}\right) } \;,
$$
\vspace{-0.1cm}
$$
 {  {M}_i\left( {f}_{ {t}},  {\rho}\right)=\frac{ {A}_{ {i}}( {\rho})}{ {B}_{ {i}}( {\rho})+\left(\frac{ {f}_{ {t}}}{\mathbf{1 0 0 c}}- {p}_{ {i}}\right)^2}  }\;,\\
$$ where
\begin{flalign*}
	\left\{ \begin{aligned}
		& A_1(\rho)\!=\!5.159\! \times \!\!10^{-5}(1\!-\!\rho)\!\left(-6.65 \!\times\!\! 10^{-5}(1\!-\!\rho)\!\!+\!0.0159\right), \\
		& A_2(\rho)=0.1925 \rho(0.1350 \rho+0.0318) \;,\\
		& A_3(\rho)=0.2251 \rho(0.1314 \rho+0.0297)\;, \\
		& A_4(\rho)=2.053 \rho(0.1717 \rho+0.0306) \;,\\
		& A_5(\rho)=0.177 \rho(0.0832 \rho+0.0213)\;, \\
		& A_6(\rho)=2.146 \rho(0.1206 \rho+0.0277)\;, \\ 
	\end{aligned}\right. &&
\end{flalign*} 
\vspace{-0.25cm}
\begin{flalign*}
	\left\{\begin{aligned}
		& B_1(\rho)=\left(-2.09 \times 10^{-4}(1-\rho)+0.05\right)^2\;, \\
		& B_2(\rho)=(0.4241 \rho+0.0998)^2 \;,\\
		& B_3(\rho)=(0.4127 \rho+0.0932)^2 \;,\\
		& B_4(\rho)=(0.5394 \rho+0.0961)^2 \;,\\
		& B_5(\rho)=(0.2615 \rho+0.0668)^2 \;,\\
		& B_6(\rho)=(0.3789 \rho+0.0871)^2 \;,\\
	\end{aligned} \right. &&
\end{flalign*}
\vspace{-0.25cm}
\begin{flalign*}
	\left\{ \begin{aligned}
		& p_1=3.96 \mathrm{~cm}^{-1}, \quad p_2=6.11 \mathrm{~cm}^{-1},\quad p_3=10.84 \mathrm{~cm}^{-1}\;,\\
		& p_4=12.68 \mathrm{~cm}^{\!-1}, \quad p_5=14.65 \mathrm{~cm}^{\!-1},\quad p_6=14.94 \mathrm{~cm}^{\!-1}.
	\end{aligned}\right. &&
\end{flalign*}

	\subsection{SNR}
\textcolor{black}{	According to the definition of SNR in the optical communication literature (e.g., \cite{ref29,ref108}), the general expression for SNR in a VLC system is given by:}

	\textcolor{black}{ \begin{equation}
		{SNR=\frac{ P_E}{\sigma_{nV}^2}=
        \frac{\left(\eta P_c\right)^2}{\sigma_{nV}^2}} ,
		\label{eq19}
	\end{equation}
	where \(P_E\) is the received electrical signal power, \(P_c\) (as defined earlier) is the transmitted optical power, \(\eta\) is the optical-to-electrical conversion efficiency, and \(\sigma_{nV}^2\) is the total noise variance in the VLC channel, which includes  shot noise (\(\sigma_{sh}^2\)) and thermal noise (\(\sigma_{th}^2\)), and can be expressed as:}
	\begin{equation}
		\sigma_{nV}^2=\sigma_{sh}^2+\sigma_{th}^2 ,
   	\label{eq20}
	\end{equation}
    where
   \begin{equation*}
		\begin{aligned}
			\sigma_{sh}^2&=2 q \eta P_c B+2 q \eta P_{bs} B,  \\
			\sigma_{th}^2&=\frac{8 \pi k_B \mathrm{~T}}{G_v} \vartheta A_R I_2 B^2+\frac{16 \pi^2 k_B \Gamma \mathrm{T}}{g_m} \vartheta^2 A_R^2 I_3 B^3,
		\end{aligned}
	\end{equation*}
	and $ P_{bs} $ represents the background light power. Other parameters are shown in Table \ref{jad1} (according to  \cite{ref108}).
 
  Also, for the THz system, the SNR is obtained as:
	\begin{equation}
		{ SNR=\frac{P_{th}}{\sigma_{nT}^2}=\frac{P_th_{pl}^2 G_{S_{ik}}\left(\varphi_{S_{ik}}, \phi_{S_{ik}}\right) G_{R_{ik}}\left(\varphi_{R_{ik}}, \phi_{R_{ik}}\right)}{\sigma_{nT}^2},} 
		\label{eq21}
	\end{equation}
where $\sigma_{nT}$ is the noise variance of THz channel.
	\begin{table}[H]
		\caption{VLC noise parameters with constant values} 
  \vspace{-0.4cm}
		\begin{center}
			\renewcommand{\arraystretch}{1.4}
			\begin{tabular}{|l|l|}
				\hline 
				$ I_2\!=\!0.562 $ \quad {\scriptsize  Noise factor \!\!(NF)}  & $g_m\!=\!30 \mathrm{mS}$ \;\quad {\scriptsize Channel conductivity}  \\
				\hline 
				$I_3\!=\!0.868 $ \quad {\scriptsize Noise  factor \!\!(NF)}  & $\Gamma=1.5$ \qquad\;\qquad {\scriptsize FET channel NF} \\
				\hline 		
				$T=300 \mathrm{~K} $\quad {\scriptsize Temperature } & $q\!=\!1.6 \!\times\! 10^{-19} \mathrm{C}$ \quad {\scriptsize Electron charge}   \\
				\hline
    $G_v\!=\!10 $ \;\;{\scriptsize Open-loop voltage gain} &$\vartheta\!=\!112 \; \frac{\mathrm{pF}}{\mathrm{cm}^2}$\quad{\scriptsize Detector capacitance  }\\
    \hline
				\multicolumn{2}{|c|}{{\scriptsize $k_B \!\!= \!\!1.38 \!\!\times \!\! 10^{-23} $$\frac{\mathrm{~J}}{\mathrm{K}}$}\;\;{\scriptsize  Boltzmann constant }} \\
				\hline
			\end{tabular}
   \vspace{-0.4cm}
			\label{jad1}
		\end{center}
	\end{table} 

	\subsection{SINR}  
\textcolor{black}{In a VLC system with both noise and interference, SINR is defined analogously to \eqref{eq19}, as described in the optical communications literature (e.g., \cite{ref77}):
	\begin{equation}
		{SINR=\frac{ P_E}{\sigma_{nV}^2+P_E^\prime}=\frac{\left(\eta P_c\right)^2}{\sigma_{nV}^2+\left(\eta ({P_{bs}+P_c^\prime})\right)^2}} ,
		\label{eq22}
	\end{equation}
where  \( P_E^\prime \) represents the total interference power received by the detector. This includes \( P_{bs} \), the power from ambient light sources such as sunlight and artificial illumination, and \( P_c^\prime \), which denotes interference originating from other VLC transmitters, such as streetlights.}

\begin{figure}[t]
		\centering
		\includegraphics[width=1\linewidth,height=0.18\textheight]{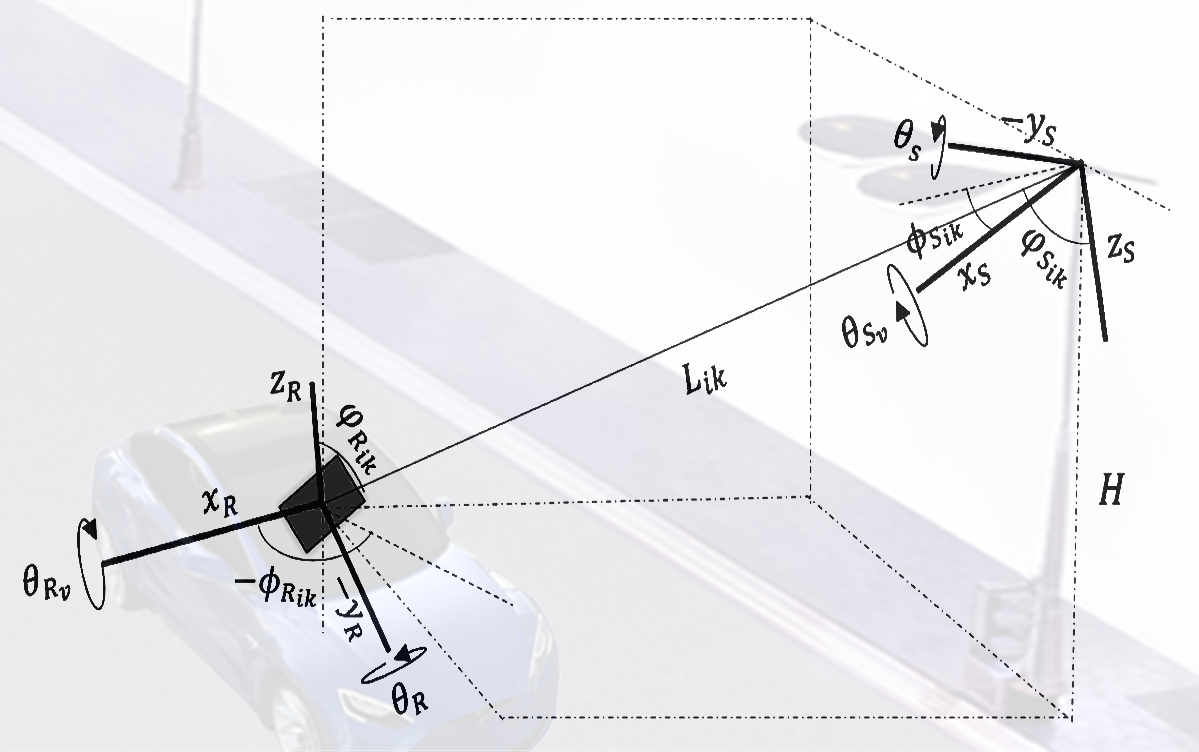}
		\vspace{-0.7cm}
		\caption{Schematic representation of the physical structure and communication  modeling between a transmitter and an interfering transmitter for calculating interference power.}
        \vspace{-0.3cm}
		\label{fig9}
	\end{figure}

To compute \( P_c^\prime \), the link length, as well as the transmitting and receiving angles, must be recalculated for the streetlights located on the opposite side of the street. We assume the streetlights on both sides of the street are symmetric in terms of their design and spacing. The only difference is first streetlight positioned slightly closer to the transverse axis by \( D_y^\prime \) to minimize interference. The concept of interference between streetlights is illustrated in Fig. \ref{fig9}.

	Based on the system modeling shown in Fig. \ref{fig9} and using the methodology discussed in the VLC system modeling section, we can calculate the interference power \( P_c^\prime \) from the light sources on the other side of the street and, in turn, derive the SINR for the VLC system.
	
For the THz system, the SINR calculation follows a similar approach. The total interference power from transmitters on the opposite side of the street, denoted as \( P_{th}^\prime \), must be considered. The SINR for the THz communication system is then given by:
	\begin{equation}
		{ SINR=\frac{P_{th}}{\sigma_{nT}^2+P_{th}^\prime}. } 
		\label{eq24}
	\end{equation}



  \begin{table*}
	\vspace{-0.6cm}
		\caption{Default parameter values}
	\begin{minipage}{0.99\textwidth}
		\begin{center}
			\renewcommand{\arraystretch}{1.3}
			\begin{tabular}{|p{3.2cm}l|l|p{2.4cm}l|p{2.47cm}l|}
				\hline 
				&\;\;VLC system &\;\;THz system &\multicolumn{2}{c|}{VLC system}& \multicolumn{2}{c|}{THz system} \\  
				\hline 
				{\scriptsize Transmitted  power} & $P_{t_V}\!\!=\!0.63 \mathrm{~W}$  & $P_{t_T}\!\!=\!0.63 \mathrm{~W}$ & {\scriptsize Lambertian coefficient}& $m=5$  & {\scriptsize  Frequency} &$f_t\!=\!{144}\mathrm{~GHz} $	 \\
				\hline 
				{\scriptsize  Receiver field of view } &$\varphi_{R_{c_V}}=\frac{\pi}{2} ^\circ$  &$\varphi_{R_{c_T}}=\frac{\pi}{2} ^\circ$  &{\scriptsize  visibility }&$Vi=50 \mathrm{~km}$  &{\scriptsize Main lobe angle} &$\varPhi_{0}=0 ^\circ$ \\
				\hline
				{\scriptsize Transmitter field of view } &$\varphi_{S_{c_V}}=\frac{\pi}{2} ^\circ $ & $\varphi_{S_{c_T}}=\frac{\pi}{2} ^\circ $ &{\scriptsize Bandwidth} &	$B=10 \mathrm{~Mhz} $ &{\scriptsize Main lobe angle} &$\varphi_{0}=0 ^\circ$  \\
				\hline
				{\scriptsize Horizontal angle of receiver  }  &$\theta_{R_V}=0 ^\circ$   & $\theta_{R_T}=0 ^\circ$  & \multicolumn{2}{c|}{{\scriptsize O/E conversion efficiency} \;\;\;\;\;$\eta=0.35 \; \frac{\mathrm{A}}{\mathrm{W}}$}& {\scriptsize Noise variance}& $\sigma_{nT}^2\!=\!27  \; \mu \mathrm{W}$ \\
				\hline 
				{\scriptsize  Horizontal angle of transmitter }  &$\theta_{S_V}=0 ^\circ $  & $\theta_{S_T}=0 ^\circ $ &{\scriptsize Transmitter area} &$A_S\!=\!0.09 ~m^{\!2}$ & {\scriptsize  Transmitter elements} & $N_S\!=\![10\;\;10]$	 \\
				\hline 
				{\scriptsize Rotation angle of receiver  }& $\theta_{R_{v_V}}\!\!=\left[ 0 \;\; 0\right] ^\circ $  & $\theta_{R_{v_T}}\!\!=\left[ 0 \;\; 0\right] ^\circ $ & \multicolumn{2}{c|}{{\scriptsize Background light power  }\;\;\; $ P_{\!b s}\!\!=\!\!0.29 \mathrm{~nW} $}& {\scriptsize Receiver elements } & $ N_R\!=\![10 \;\; 10] $ \\
				\hline 
				{\scriptsize Rotation angle of transmitter  }& $\theta_{S_{v_V}}\!\!=\left[ 0 \;\; 0\right] ^\circ $  & $\theta_{S_{v_T}}\!\!=\left[ 0 \;\; 0\right] ^\circ $ & 				\multicolumn{3}{c}{{\scriptsize Side angle of light rods and  transverse axis }}&\multicolumn{1}{l|}{ $\theta_{in}\!=\!\left[ 0 \;\; 0 \right] \,^\circ$}  \\
				\hline 
				{\scriptsize Angle of each light and its rod } &
				$\theta_{o_V}=\left[0 \;\; 0 \right] ^\circ $  & $\theta_{o_T}=\left[ 0 \;\; 0\right] ^\circ $  & \multicolumn{3}{l}{{\scriptsize \!The distance difference of  the poles of both 	street side from the origin\!\!}}&\multicolumn{1}{l|}{  $ D_y^{\prime}=\frac{D_y}{2} $ } \\
				\hline 
			\end{tabular}
   
   \vspace{0.1cm}
   \begin{tabular}{|p{5.22cm}|p{3.5cm}|p{3.3cm}|p{4.37cm}|}
   \hline
    		{\scriptsize Width of  the street ( \!each direction)  } \, $ W\!D_l\!\!=\!10\mathrm{~m} $ &
				{{\scriptsize Height of   light poles } \,  $ H_{\!t}\!\!=\!10\mathrm{~m} $} &{\scriptsize Height of  the car}  \,$ H_{r}\!\!=\!1.5\mathrm{~m} $& {{\scriptsize Distance between  light poles } \,  $ D_{\!y}\!\!=\!30\mathrm{~m} $} \\
				\hline 
    \multicolumn{4}{|l|}{Receiver position parameters on the car roof: \quad $ d_{rxf}=\!d_{rxb}\!=\! 0.75 \mathrm{~m} ,\quad d_{ryf}\!=\!d_{ryb}\!=\!0.75 \mathrm{~m} ,\quad  d_{cyb}\!=\!2 \mathrm{~m} ,\!\quad d_{cyf}\!=\!3 \mathrm{~m},\quad d_{cxr}\!=\!d_{cxl}\!=\!1 \mathrm{~m} $}\\
    \hline
   \end{tabular}
			\label{jad2}
		\end{center}
	\end{minipage}
 \vspace{-0.4cm}
 \end{table*}

\begin{figure*}[!t]
		\subfloat[\label{fig5_1a}]{\includegraphics[width=0.25\linewidth,height=0.16\textheight]{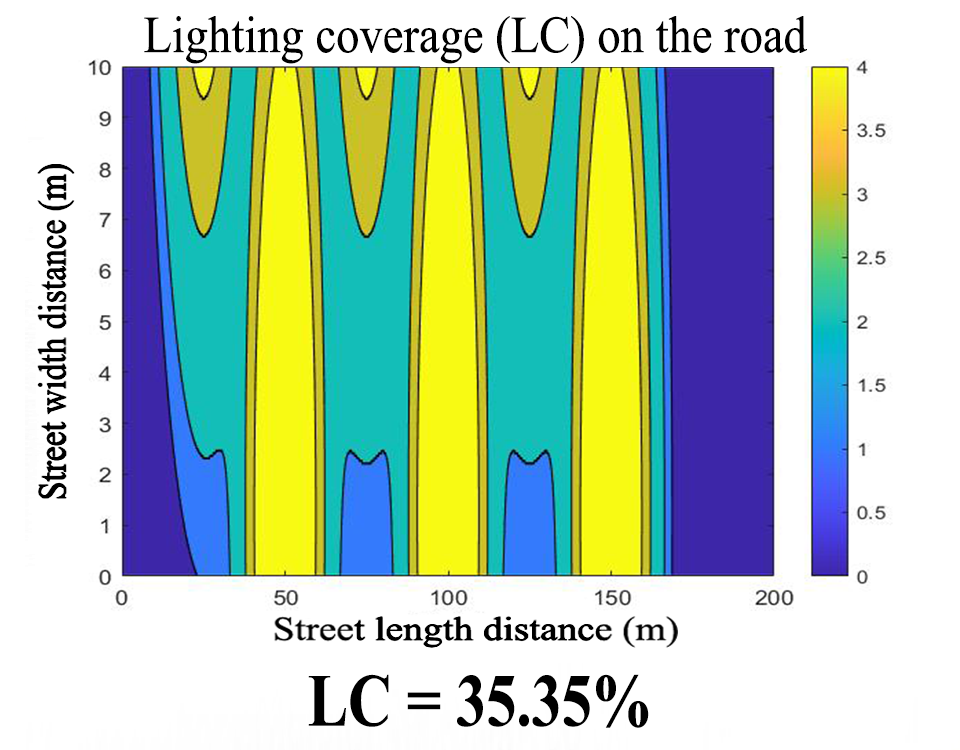}}%
		\subfloat[\label{fig5_1b}]{\includegraphics[width=0.25\linewidth,height=0.16\textheight]{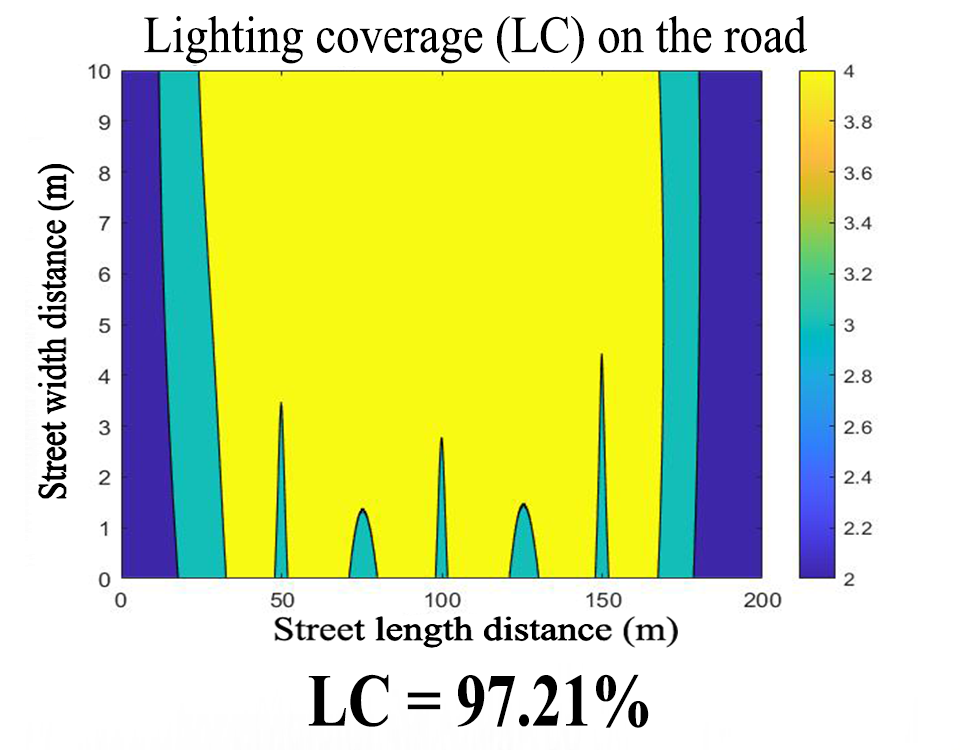}}%
		\subfloat[ \label{fig5_1c}]{\includegraphics[width=0.25\linewidth,height=0.16\textheight]{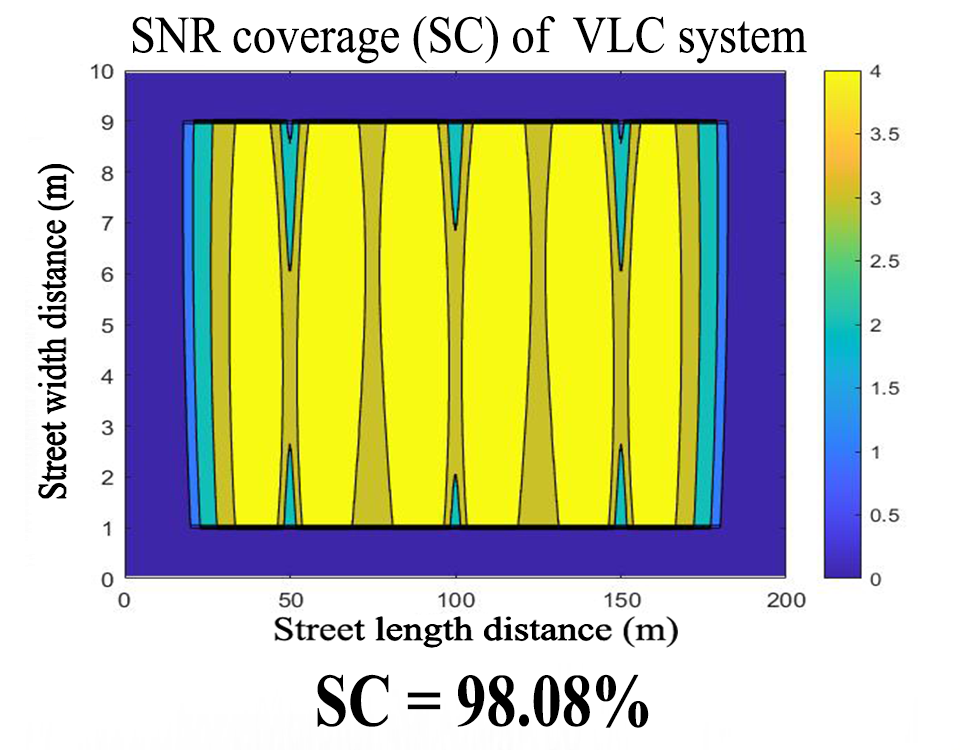}}%
		\subfloat[\label{fig5_1d}]{\includegraphics[width=0.25\linewidth,height=0.16\textheight]{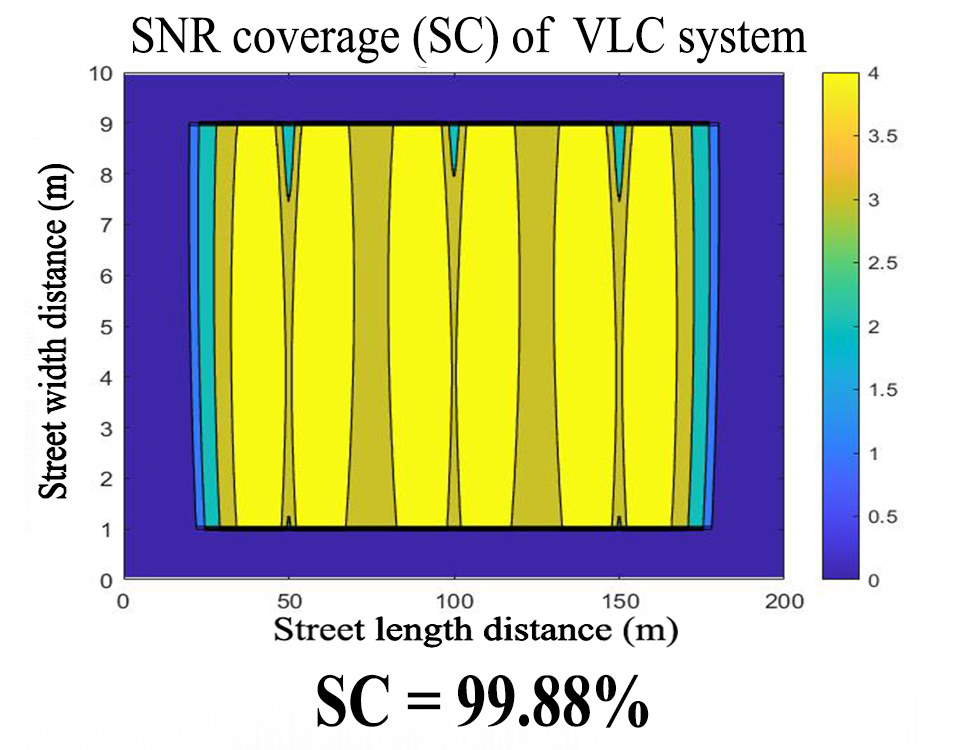}}       
		\vspace{-0.2cm}
		\subfloat[ \label{fig5_1e}]{\includegraphics[width=0.25\linewidth,height=0.16\textheight]{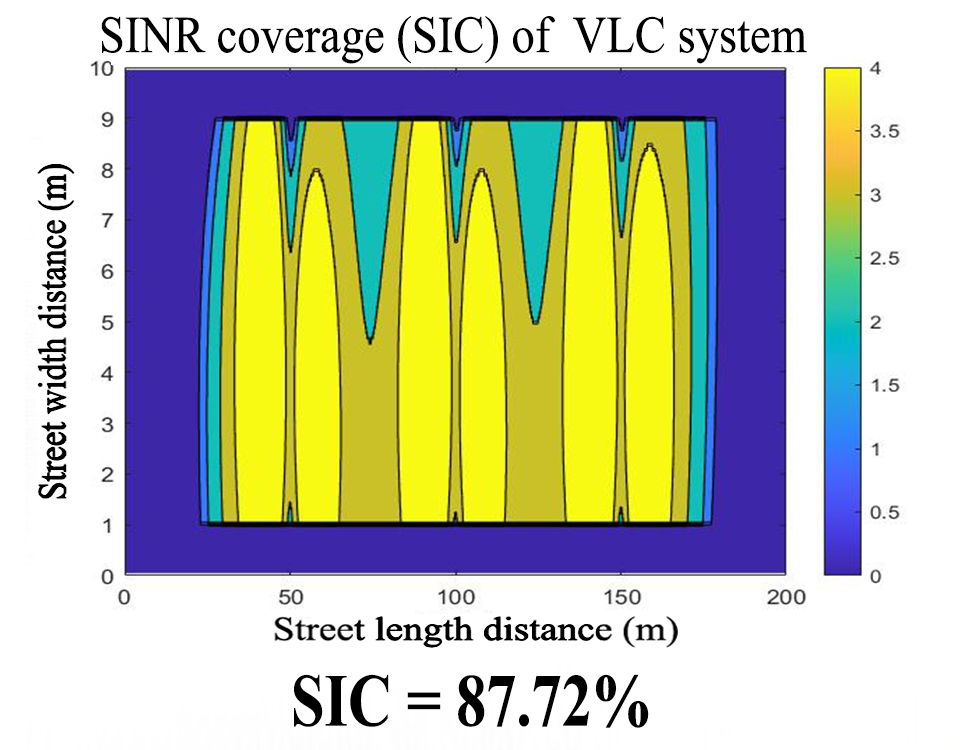}}%
		\subfloat[ \label{fig5_1f}]{\includegraphics[width=0.25\linewidth,height=0.16\textheight]{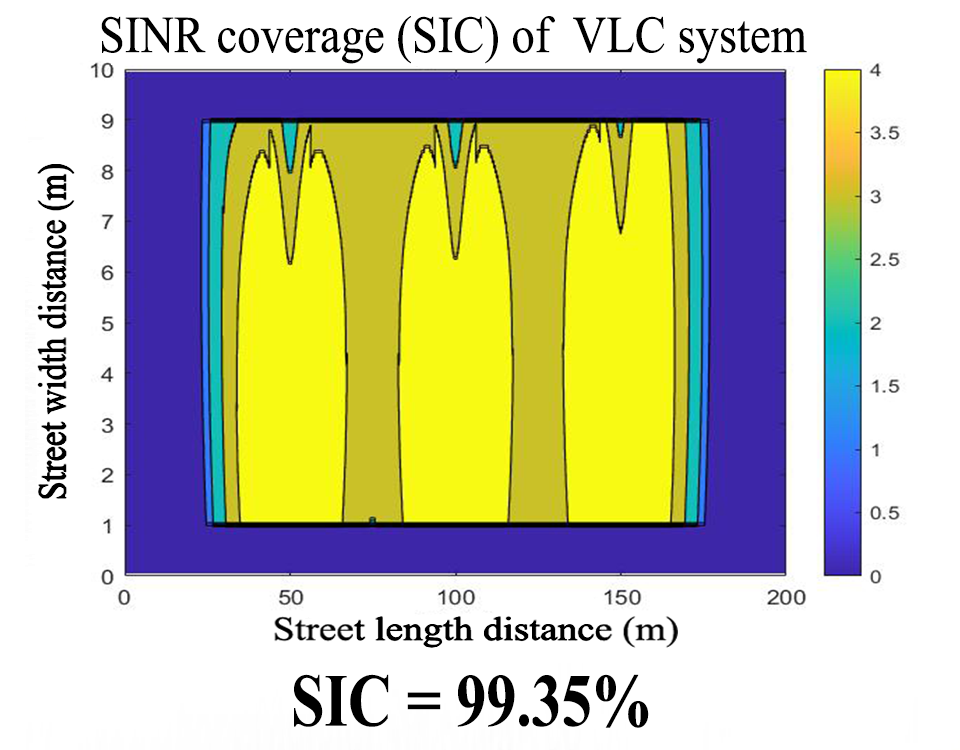}}%
		\subfloat[ \label{fig5_1i}]{\includegraphics[width=0.25\linewidth,height=0.16\textheight]{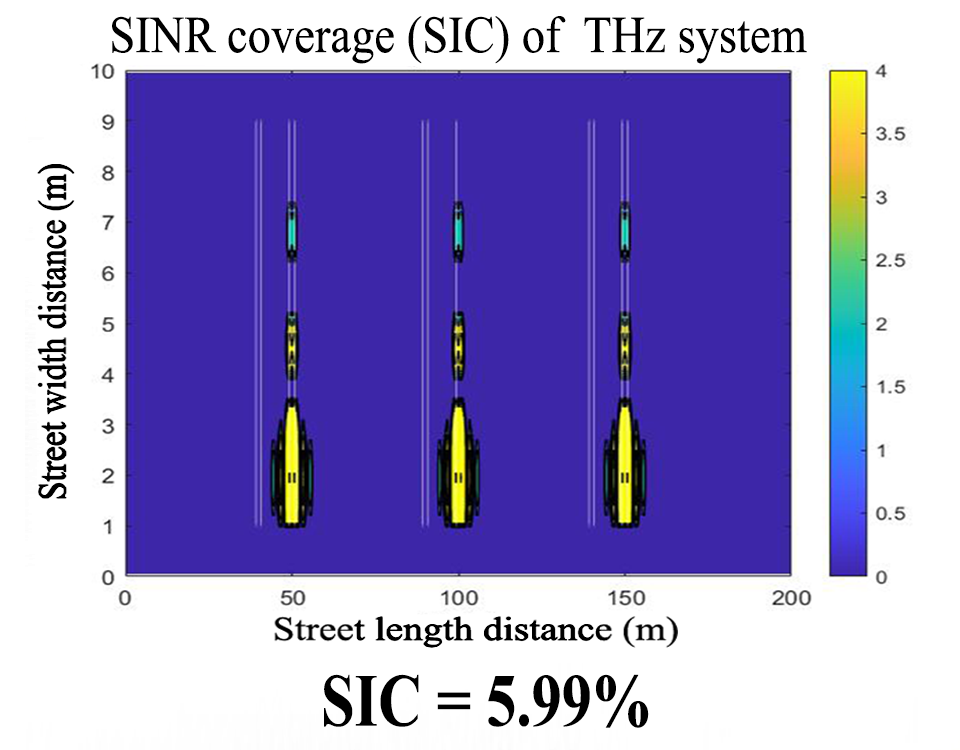}}%
		\subfloat[\label{fig5_1j}]{\includegraphics[width=0.25\linewidth,height=0.16\textheight]{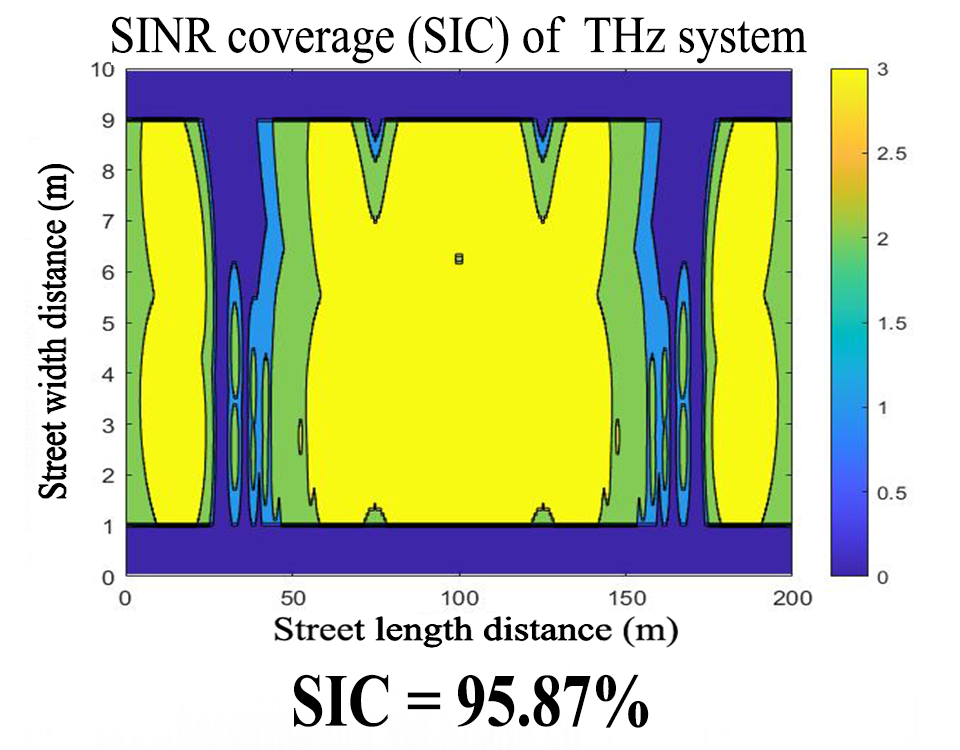}}%
		\caption{\textcolor{black}{Display of coverage on the street surface according to (a)lighting coverage (LC), (b) LC with optimal parameters, (c) SNR coverage (SC), (d) SC with optimal parameters, (e) SINR coverage (SIC) of VLC, (f) SIC of VLC  with optimal parameters, (g) SIC of THz, (h) SIC of  THz with optimal parameters.}}
		\label{fig:fig5_1}
		\vspace{-0.5cm}
	\end{figure*}
    
 \subsection{Outage Probability}
	We know that the received signal by the vehicle is:
	$$
	y(t)=h(t) x(t)+n(t),
	$$
	where $ n(t) $ is a white Gaussian noise with zero mean and variance of $\sigma_n$. Therefore, the probability distribution and cumulative distribution of $ y(t) $ are as follows, respectively:
	\begin{equation}
		{ f_y(y)=\frac{1}{\sigma \sqrt{2 \pi}} \exp \left\{-\frac{(y-\eta P_c)^2}{2 \sigma_n^2}\right\},} 
		\label{eq4_66}
	\end{equation}
	\begin{equation}
		{ F_y(y)=Q\left(\frac{y-\eta P_c}{\sigma_n}\right), } 
		\label{eq4_67}
	\end{equation}
	where  $Q(y)$ is the well-known Q-function.
	
	The above equations is true for both VLC and THz systems, so according to  \cite{ref111},  the outage probability of the hybrid system for GS and PSC, respectively, are as follows:
	\begin{equation}
		{	P_{\text{\tiny {PSC}}}^{\text{{\tiny OP}}}\!\left(y_{th}\right)\!\!=\!\!P\left[y_{V}\!<y_{th}, y_{T}\!<\gamma_{th}\right]\!\!=\!\!F_{y_{V}}\!\!\left(y_{th}\right) \!F_{y_{T}}\!\!\left(y_{th}\right) },
		\label{eq26}
	\end{equation}
 { \begin{equation}
  	\label{eq27}
			\!\!\!\!P_{\text{\tiny {GS}}}^{\text{{\tiny OP}}}(\!\gamma_{th_l}, \gamma_{th_h}\!)\! =\!  F_{\gamma_V}\! (\gamma_{th_l}\! ) F_{\gamma_T}\! (\gamma_{th_l}) \!\! \times\! [1\! +\!F_{\gamma_T}\! (\gamma_{th_l})\times
    \vspace{-0.15cm}
   \end{equation}
   \begin{equation*}
   \begin{aligned}
   &\quad(F_{\gamma_V}(\! \gamma_{th_h}\! )\!\!-\!\!F_{\gamma_V}\!(\!\gamma_{th_l}))\!+\!F_{\gamma_V}\! (\gamma_{th_l})(F_{\gamma_T}\! (\! \gamma_{th_h}\! )\!-\!\!F_{\! \gamma_T}\! (\gamma_{th_l}\!) )\\
			&\quad\!+\!(\! F_{\! \gamma_V}\! (\gamma_{th_h}\!)\!\!-\!\!F_{\gamma_V}\!(\!\gamma_{th_l}\!))(\! F_{\! \gamma_V}(\! \gamma_{th_h}\!)\!\!-\!\!F_{\! \gamma_V}\!(\!\gamma_{th_l}\!))	(F_{\! \gamma_{co}}\!(\gamma_{th_h}\!))].
		\end{aligned}
   \end{equation*}}

   \section{Simulation Results and  Analysis}

\begin{figure*}[!t]
		\vspace{-0.5cm}
		\subfloat[\label{fig5_2a}]{\includegraphics[width=0.25\linewidth,height=0.16\textheight]{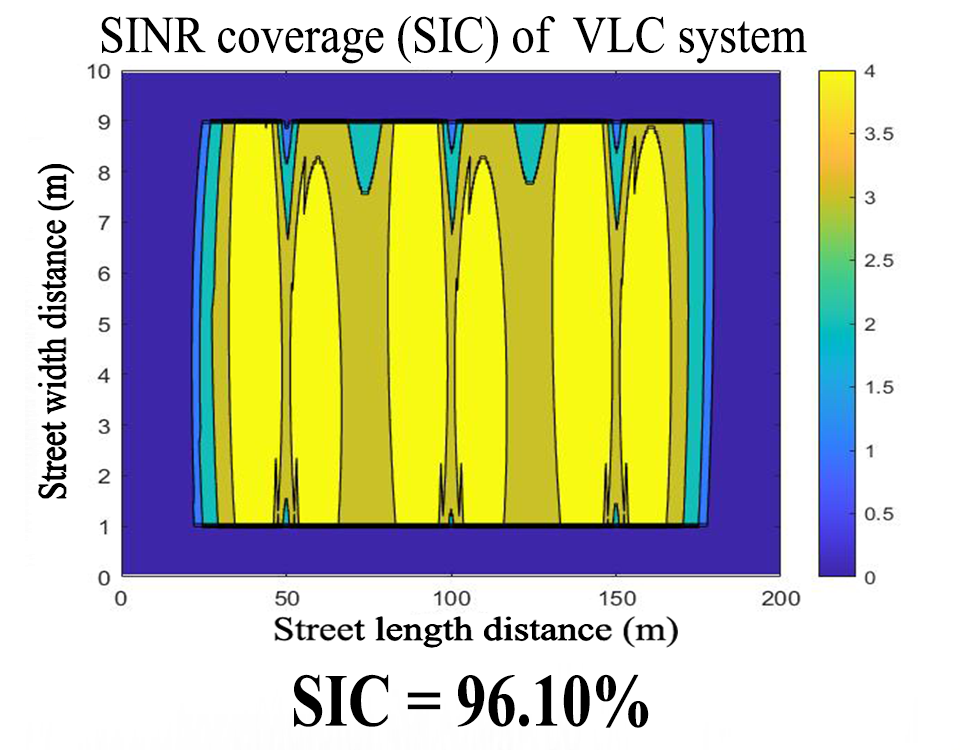}}%
		\subfloat[\label{fig5_2b}]{\includegraphics[width=0.25\linewidth,height=0.16\textheight]{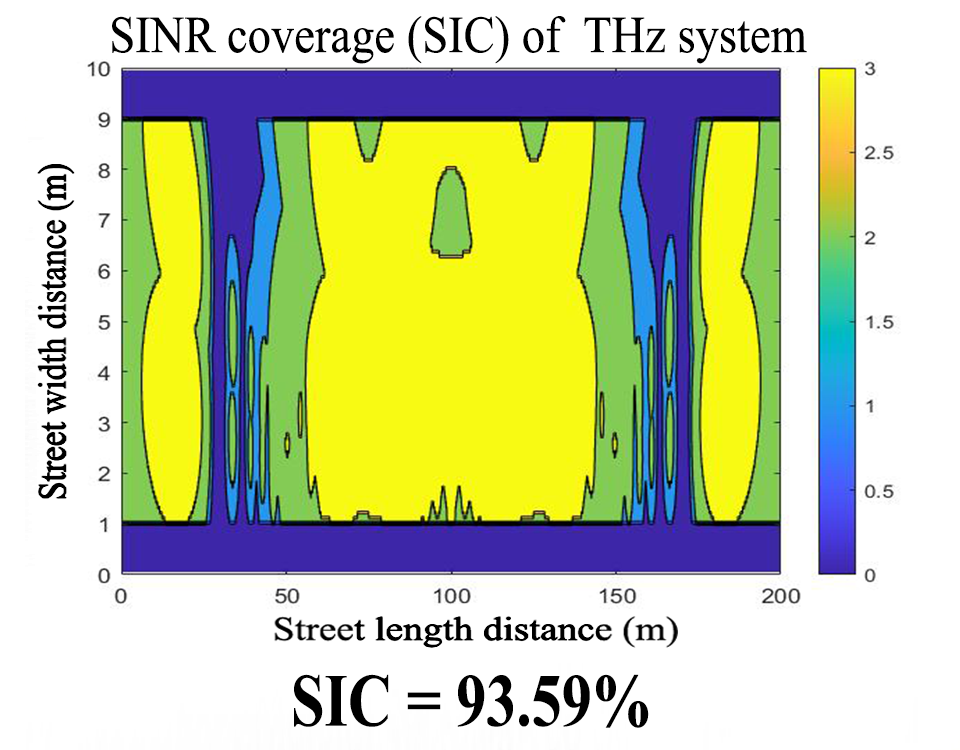}}%
		\subfloat[\label{fig5_2c}]{\includegraphics[width=0.25\linewidth,height=0.16\textheight]{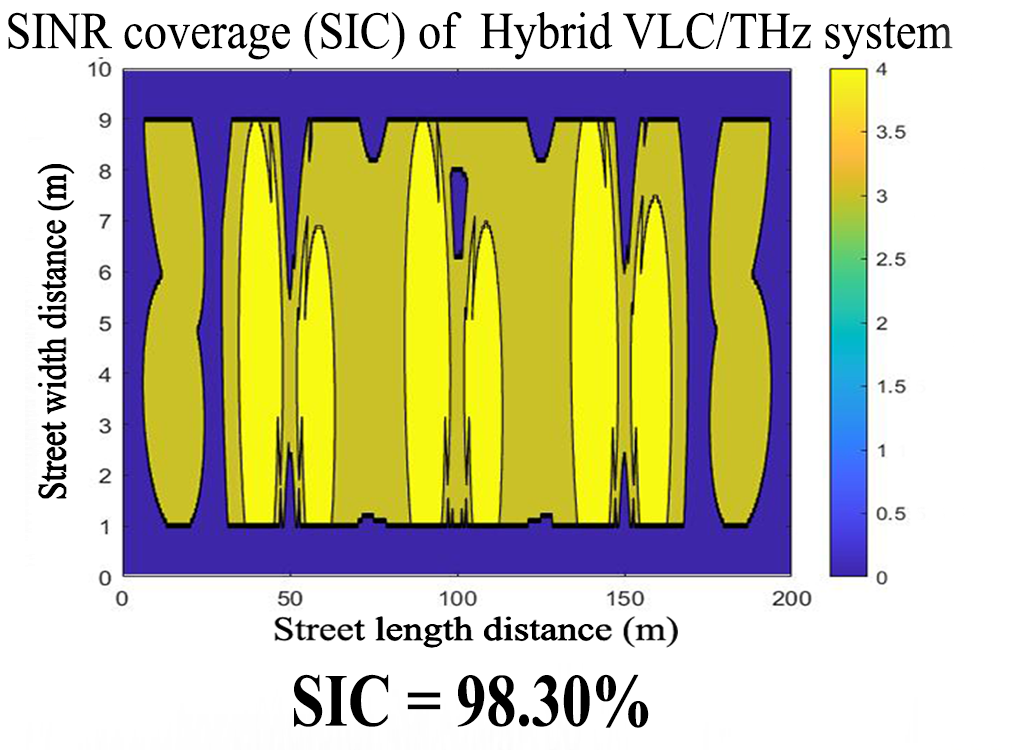}}%
		\subfloat[ \label{fig5_2d}]{\includegraphics[width=0.25\linewidth,height=0.16\textheight]{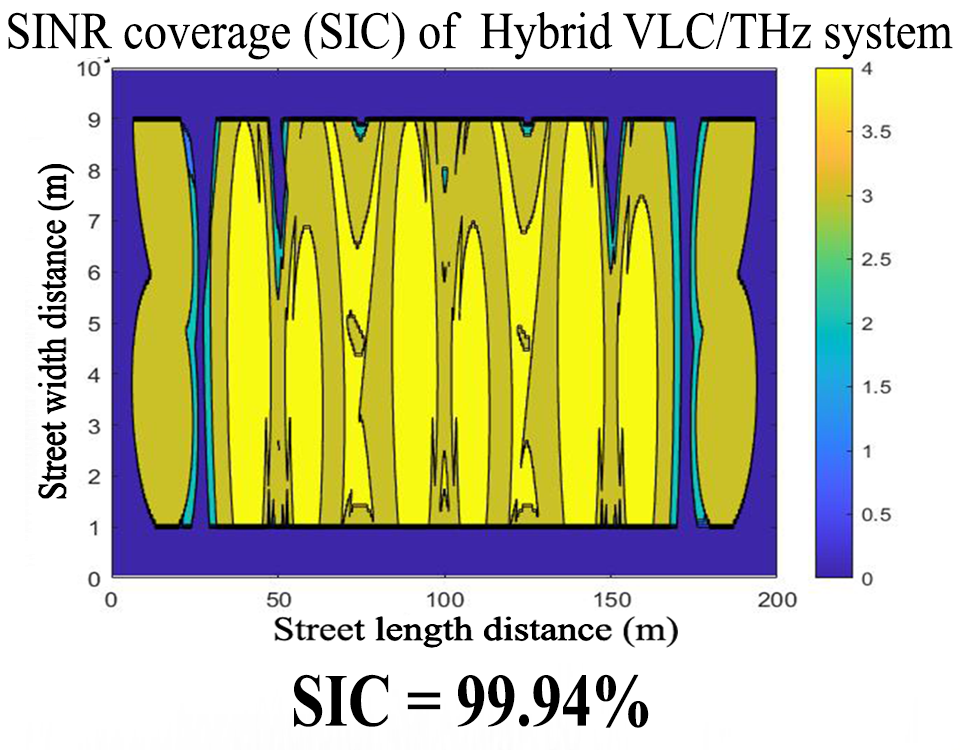}}       
		\vspace{-0.2cm}
		\subfloat[ \label{fig5_2e}]{\includegraphics[width=0.25\linewidth,height=0.16\textheight]{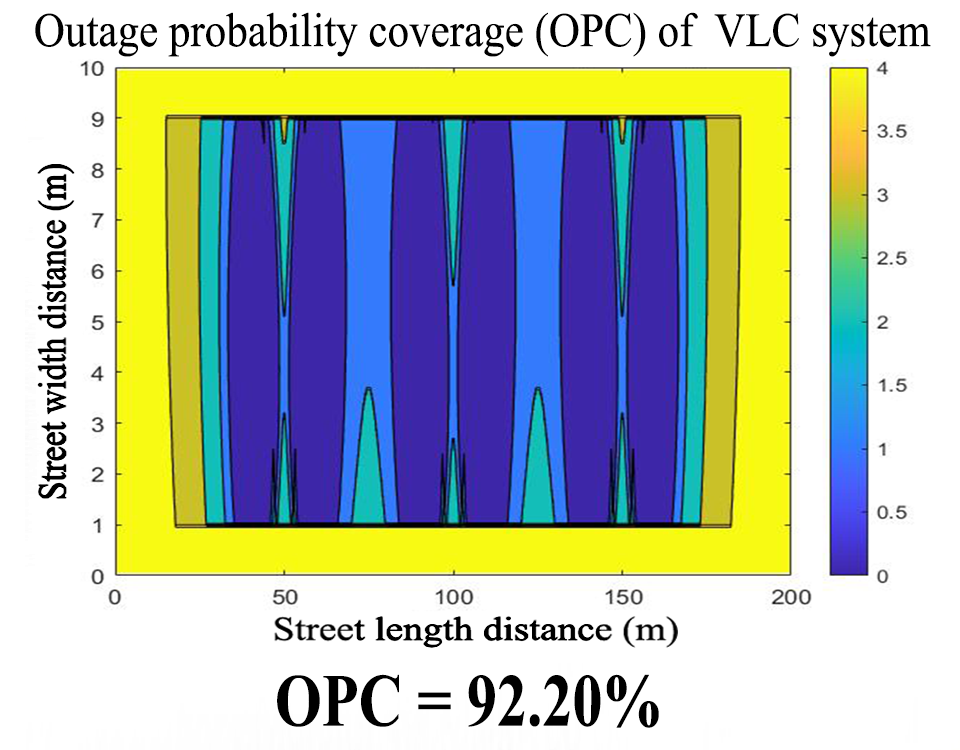}}%
		\subfloat[\label{fig5_2f}]{\includegraphics[width=0.25\linewidth,height=0.16\textheight]{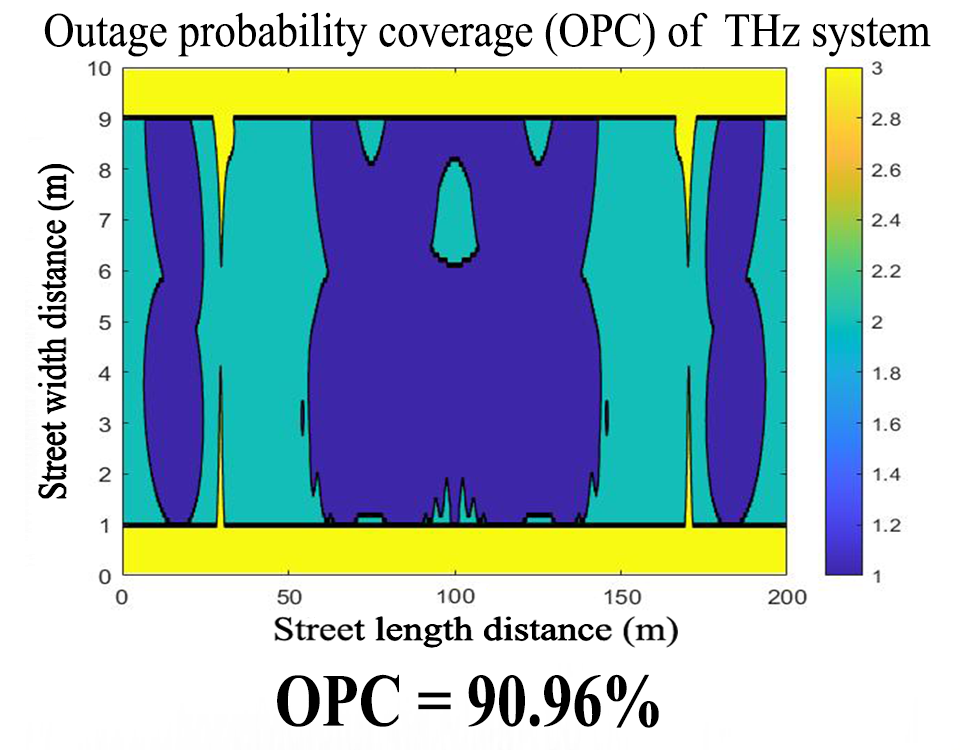}}%
		\subfloat[\label{fig5_2g}]{\includegraphics[width=0.25\linewidth,height=0.16\textheight]{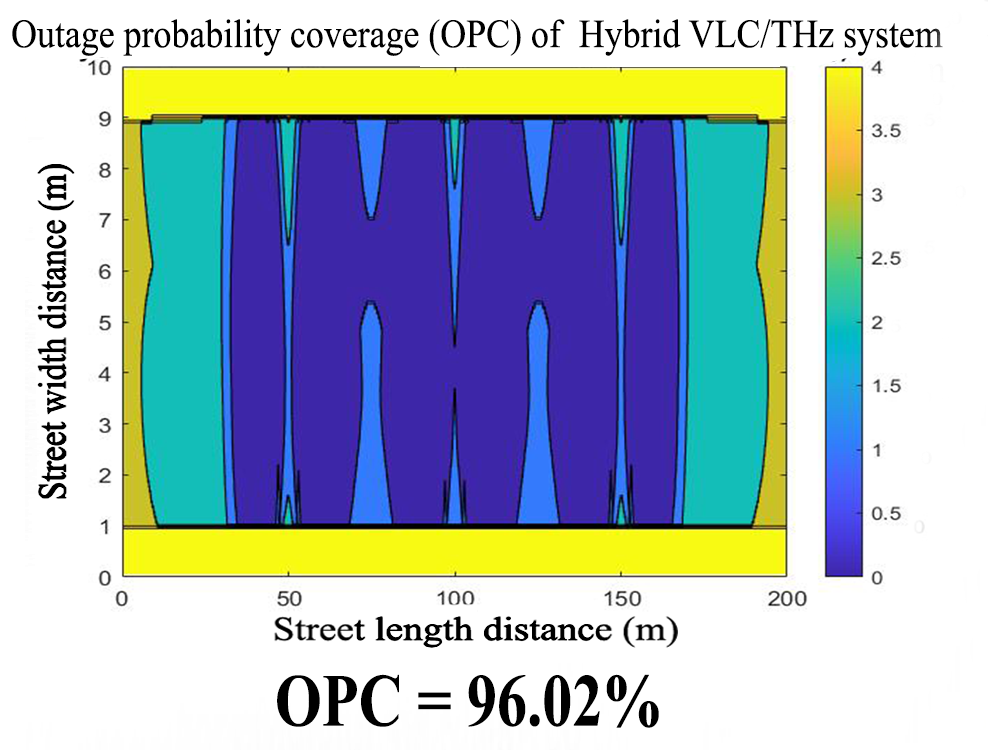}}%
		\subfloat[ \label{fig5_2h}]{\includegraphics[width=0.25\linewidth,height=0.16\textheight]{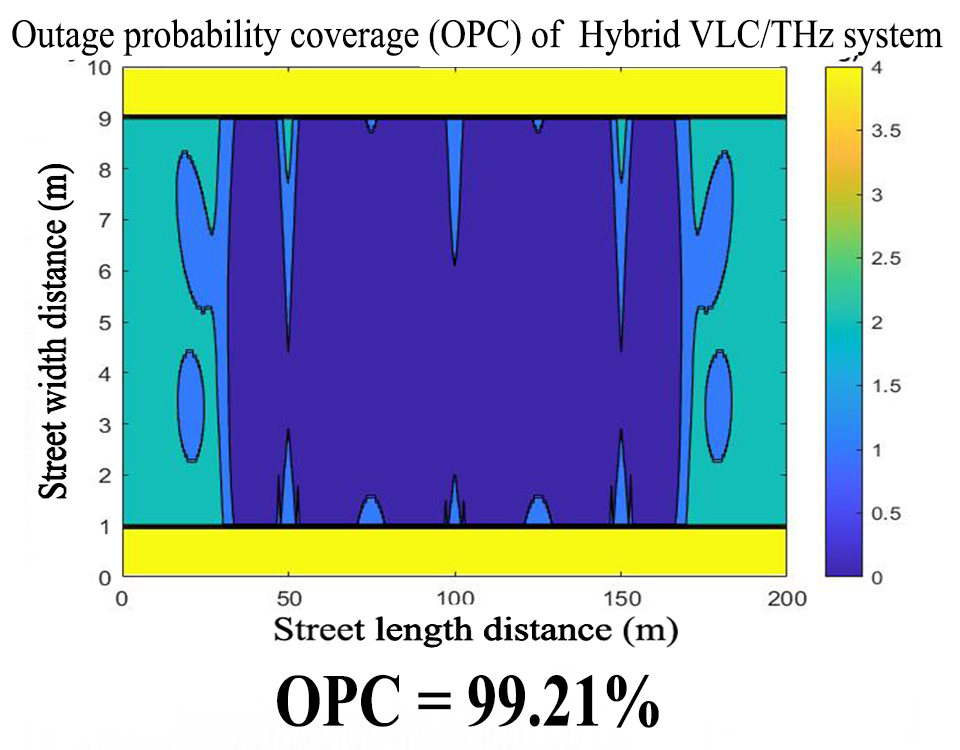}}%
		\caption{\textcolor{black}{Display of coverage on the street surface  with chosen hybrid parameters according to (a) SIC of VLC, (b)  SIC of THz, (c) SIC of GS,  (d) SIC of PSC, (e) outage probability coverage (OPC) of VLC, (f) OPC of THz,  (g) OPC of GS, (h) OPC of PSC.}}
		\label{fig:fig5_2}
		\vspace{-0.4cm}
	\end{figure*}

In this section, we analyze the lighting coverage (LC) on the street surface and extend this analysis to vehicular communication applications using three key performance metrics: SNR, SINR, and outage probability. At each stage, we employ a parameter optimization technique known as the ``grid search" to identify optimal parameters for each scenario. Table \ref{jad2} presents the default values of the system parameters. Figs. \ref{fig:fig5_1} and \ref{fig:fig5_2} depict the coverage achieved on the street surface according to each of these performance metrics. In these figures, coverage quality is represented at four distinct levels, indicated by different colors in the color bar. The four threshold levels in Figs. \ref{fig5_1a} and \ref{fig5_1b} are set at $[4 \times 10^{-8}, 1 \times 10^{-7}, 6 \times 10^{-7}, 2 \times 10^{-6}]\, \text{W}$, while in Figs. \ref{fig5_1c} to \ref{fig5_1j} and Figs. \ref{fig5_2a} to \ref{fig5_2d}, the thresholds are $[-1, 1, 5, 10]\, \text{dB}$, and in Figs. \ref{fig5_2e} to \ref{fig5_2h}, the thresholds are $[10^{-7}, 10^{-5}, 10^{-3}, 10^{-2}]$. Additionally, for calculating coverage percentage, we set the following threshold values: for LC, $2 \times 10^{-6}\, \text{W}$, for SNR and SINR, $5\, \text{dB}$, and for outage probability, $10^{-6}$. Moreover, as vehicles cannot pass through the longitudinal borders of the street, the receivers do not receive any power within the distances from 0 to 1 meter and from 9 to 10 meters. 

We begin by dividing the street surface into a series of small grids. For each grid, we calculate the received power based on the modeling described in Section III. The street structure under consideration has a width ($WD_l$) of 10 meters and a pole height ($H_t$) of 10 meters. According to city standards, the longitudinal distance between streetlights is typically set at approximately 30 meters. However, in order to reduce infrastructure costs, we aim to extend this distance.
	\begin{table}[t]
		\vspace{-0.01cm}
		\caption{  Optimal Parameter Values}
		\vspace{-0.35cm}
		\begin{center}
			\renewcommand{\arraystretch}{1.3}
			\begin{tabular}{|c|c|c|}
				\hline
				{LC} & SC & SIC \\
				\hline 
				$m=6$ & $m=6$ & $m=6$ \\
				\hline 
				$D_y=50\,\mathrm{~m}$ & $D_y=50\,\mathrm{~m}$ & $D_y=50\,\mathrm{~m}$ \\
				\hline 
				$D_y^{\prime}=8\,\mathrm{~m}$ & $D_y^{\prime}=8\,\mathrm{~m}$ & $D_y^{\prime}=8\,\mathrm{~m}$ \\
				\hline 
				$\theta_{o}=[79\;\;79]\,^\circ$ & $\theta_{o}=[79\;\;79]\,^\circ$ & $\theta_{o}=[79\;\;79]\,^\circ$ \\
				\hline 
				$\theta_{in}=[0\;\;0]\,^\circ$ & $\theta_{in}=[0\;\;0]\,^\circ$ & $\theta_{in}=[0\;\;0]\,^\circ$ \\
				\hline 
				$\theta_S=12\,^\circ$ & $\theta_S=12\,^\circ$ & $\theta_S=12\,^\circ$ \\
				\hline 
				$\theta_{S_v}=[56\;\;56]\,^\circ$ & $\theta_{S_v}=[56\;\;56]\,^\circ$ & $\theta_{S_v}=[56\;\;56]\,^\circ$ \\
				\hline 
				$\phi_{S_c}=[90\;\;90]\,^\circ$ & $\phi_{S_c}=[90\;\;90]\,^\circ$ & $\phi_{S_c}=[90\;\;90]\,^\circ$ \\
				\hline 
				& $\phi_{R_c}=[90\;\;90]\,^\circ$ & $\phi_{R_c}=[71\;\;71]\,^\circ$ \\
				\hline 
				& $\theta_R=24\,^\circ$ & $\theta_R=27\,^\circ$ \\
				\hline 
				& $\theta_{R_v}=[28\;\;28]\,^\circ$ & $\theta_{R_v}=[46\;\;46]\,^\circ$ \\
            \hline
			  \vspace{-0.05cm}	{LC=97.21\%} & {SC=99.88\%}  & {SIC=95.73\%} \\
           \textcolor{black}{ (Fig. \ref{fig5_1b})}&  \textcolor{black}{ (Fig. \ref{fig5_1d}) }& \\
				\hline 
			\end{tabular}
			\label{jad3}
		\end{center}
                \vspace{-0.85cm}
	\end{table}
\begin{table}[t]
		\vspace{-0.01cm}
		\caption{Optimal parameters obtained for the considered simulations.}
		\vspace{-0.4cm}
		\begin{center}
			\renewcommand{\arraystretch}{1.3}
			\begin{tabular}{|c|c|c|c|}
				\hline
				\multirow{2}{*}{Single VLC} & \multirow{2}{*}{Single THz} & \multicolumn{2}{c|}{{Hybrid System}} \\
				\cline{3-4} 	
				&&Hybrid VLC & Hybrid THz \\
				\hline 
				$m=6$ & $N_R\!=\![10 \; 10]$ & $m=6$ & $N_R=[10 \; 10]$ \\
				\hline 
				$D_y=50\; m$ & $N_S=[9\;\;9]$ & $D_y=50\; m$ & $N_S=[9\;\;9]$  \\
				\hline 
				$D_y^{\prime}=25\; m$ & $\theta_{o}\!\!=\!\![77 \;\; 77]\,^\circ$ & $D_y^{\prime}=10\; m$ &$\theta_{o}\!\!=\!\![67 \;\; 67]\,^\circ$ \\
				\hline 
				$\theta_{o}\!\!=\!\![73 \;\; 73]\,^\circ$ & $\theta_S=34\,^\circ$ & $\theta_{o}\!\!=\!\![84 \;\; 84]\,^\circ$ & $\theta_S=37\,^\circ$ \\
				\hline 
				$\theta_{in}=[0 \;\; 0]\,^\circ$ & $\theta_R=6\,^\circ$ & $\theta_{in}=[0 \;\; 0]\,^\circ$  & $\theta_R=6.5\,^\circ$ \\
				\hline 
				$\theta_S=9\,^\circ$ & $\!\!\theta_{S_v}\!\!\!=\![\!-\!6 \; \!-\!6]^\circ\!\!$ & $\theta_S=10\,^\circ$ &$\!\!\theta\!_{S\!_v}\!\!\!=\![\!-\!6.\!7 \; \!-\!\!6.\!7]^\circ$\!\!\! \\
				\hline 
				{\scriptsize $\theta_{S_v}\!\!=\!\![48 \;\; 48]\,^\circ$ }& {\scriptsize $\theta_{R_v}\!\!\!=\!\![77 \;\; 77]^\circ$} &{\scriptsize  $\theta_{S_v}\!\!=\!\![55 \;\; 55]\,^\circ$ }& {\scriptsize $\theta_{R_v}\!\!=\!\![77 \;\; 77]\,^\circ$}\\
				\hline
				$\theta_R=20\,^\circ$ & $\varPhi_{0}=0\,^\circ$ & $\theta_R=27\,^\circ$ & $\varPhi_{0}=0\,^\circ$ \\
				\hline
				{\scriptsize 	$\theta_{R_v}\!\!=\!\![43 \;\; 43]\,^\circ$ }& $\varphi_{0}=4\,^\circ$ & {\scriptsize $\theta_{R_v}\!\!=\!\![46 \;\; 46]\,^\circ$} & $\varphi_{0}=4.1\,^\circ$ \\
				\hline
				{\scriptsize $\phi_{S_c}\!\!=\!\![90 \;\; 90]\,^\circ$ }& $\varphi_{R_c}=90\,^\circ$ & {\scriptsize $\phi_{S_c}\!\!=\!\![90 \;\; 90]\,^\circ$} & $\varphi_{R_c}=90\,^\circ$ \\
				\hline
				{\scriptsize 	$\phi_{R_c}\!\!=\!\![74 \;\; 74]\,^\circ$} & $\varphi_{S_c}=90\,^\circ$ & {\scriptsize $\phi_{R_c}\!\!=\!\![72 \;\; 72]\,^\circ$} & $\varphi_{S_c}=90\,^\circ$ \\
				\hline
				{SIC=99.35\%} & {SIC=95.87\%} & {SIC=96.10\%} & {SIC=93.59\%} \\
           \textcolor{black}{ (Fig. \ref{fig5_1f})}&  \textcolor{black}{ (Fig. \ref{fig5_1j}) }& \textcolor{black}{ (Fig. \ref{fig5_2a})} &  \textcolor{black}{ (Fig. \ref{fig5_2a})} \\
				\hline
				\multicolumn{2}{c}{} & \multicolumn{2}{|c|}{$ {\mathrm{LC} =96.90\%}$} \\
				\multicolumn{2}{c}{} & \multicolumn{2}{|c|}{$ {\mathrm{SIC \; of \; GS }\!=\!98.30\%}$  \textcolor{black}{(Fig.  \ref{fig5_2c})}} \\
				\multicolumn{2}{c}{} & \multicolumn{2}{|c|}{$ {\mathrm{SIC \; of \; PSC}\!=\!99.94\%}$  \textcolor{black}{(Fig.  \ref{fig5_2c})}} \\
				\cline{3-4} 	
			\end{tabular}
			\vspace{-0.005cm}
			\label{jad4}
		\end{center}
	\end{table}
	\begin{table}[!h]
		\vspace{-0.45cm}
		\caption{Different coverage values under various conditions}
		\vspace{-0.4cm}
		\begin{center}
			\renewcommand{\arraystretch}{1.3}
			\begin{tabular}{|c|c|c|c|c|c|}
				\hline
				\multirow{2}{*}{{\scriptsize Visibility} } & \multirow{2}{*}{{$\!\!\!\text{\scriptsize BG Power (\!W\!)}\!\!$}} & \multirow{2}{*}{{$\!\!\!\text{\scriptsize  VLC Coverage}\!\!$}} & \multicolumn{3}{c|}{Hybrid  Coverage} \\
				\cline{4-6} 	
				&&&NPSC&GS&PSC \\
	\hline 
				$25 \mathrm{~km}$ & $\!\!2.9 \!\times \!{10}^{\!-10\!} \!\! $ & $ 96.07 \% $&$99.41$&$98.29 \% $ & $ 99.94 \% $ \\
				\hline 
				$20 \mathrm{~km}$ & $2.9\! \times\! {10}^{-8}  $ & $ 95.98 \% $&$99.41$&$98.27 \% $ & $ 99.94 \% $ \\
				\hline
				$10 \mathrm{~km}$ & $2.9\! \times\! {10}^{-7}  $ & $ 95.15 \% $&$99.41$&$98.14\% $ & $ 99.94 \% $ \\
				\hline 
				$1 \mathrm{~km}$ & $2.9\! \times\! {10}^{-7}  $ & $ 93.98\% $&$99.41$&$98.00 \% $ & $ 99.94 \% $ \\		
				\hline 
				$0.5 \mathrm{~km}$ & $2.9 \!\times \!{10}^{-7}  $ & $91.81\% $&$99.41$&$97.75 \% $ & $ 99.87\% $ \\
				\hline 
				$0.5 \mathrm{~km}$ & $2.9\! \times\! {10}^{-6}  $ & $ 69.42 \% $&$98.64$&$94.84 \% $ & $ 99.12 \% $ \\
				\hline
			\end{tabular}
			\vspace{-0.7cm}
			\label{jad5}
		\end{center}
	\end{table}
    
    Analyzing the results, we concluded that full coverage can be maintained up to a distance of about 40 meters between streetlights. As a result, we extend this  further to 50 meters and attempt to optimize the system parameters for maximum coverage. Using the default values shown in Table \ref{jad2}, we obtain the coverage results illustrated in Fig. \ref{fig5_1a}, showing a coverage percentage of $35.35\%$. Applying our optimization method, we determine the optimal  values, listed in the first column of Table \ref{jad3}, which leads to the results in Fig. \ref{fig5_1b}, achieving an impressive coverage  of $97.21\%$.

 Then, we proceed  to design  the communication system. The objective is to optimize parameter values that ensure highly reliable communication coverage under a VLC system for the street, based on the LC requirements examined in the previous section. In this phase, we assess two key metrics:  SNR and  SINR for a vehicle moving along the street. It is important to account for the receiver angles in this context. Initially, we set the receiver angles to the same of the transmitter angles specified in Table \ref{jad3}, resulting in Fig. \ref{fig5_1c}. Subsequently, we implement our optimization method to derive optimal values for the receiver angles, which are detailed in the second column of Table \ref{jad3}. This optimization allows us to achieve an SNR coverage (SC) of $99.88\%$, while simultaneously ensuring an LC of over $97\%$, as illustrated in Fig. \ref{fig5_1d}.
 Next, we examine communication coverage in terms of SINR, starting from the updated receiver angles in the second column of Table \ref{jad3}. This analysis yields Fig. \ref{fig5_1e}, which indicates that the SINR coverage (SIC) for the vehicle stands at $87.72\%$. By iteratively optimizing these values while considering LC, we arrive at a SIC of $95.73\%$, as shown in the third column of Table \ref{jad3}. However, when we optimize without factoring in LC, we achieve a higher coverage of $99.35\%$, represented in Fig. \ref{fig5_1f}. These results are summarized in the first column of Table \ref{jad4}. This configuration offers the highest level of SIC for the VLC system when LC is not taken into account. However, given the importance of LC considerations, we opt for a trade-off between LC and SIC, resulting in the values presented in the third column of Table \ref{jad4}, which yield an LC of $96.90\%$ and an SIC of $96.10\%$. These optimized settings  are used in our hybrid system, illustrated in Fig. \ref{fig5_2a}.

	\begin{figure*}[!t]
		\subfloat[ \label{fig5_3a}]{\includegraphics[width=0.25\linewidth,height=0.14\textheight]{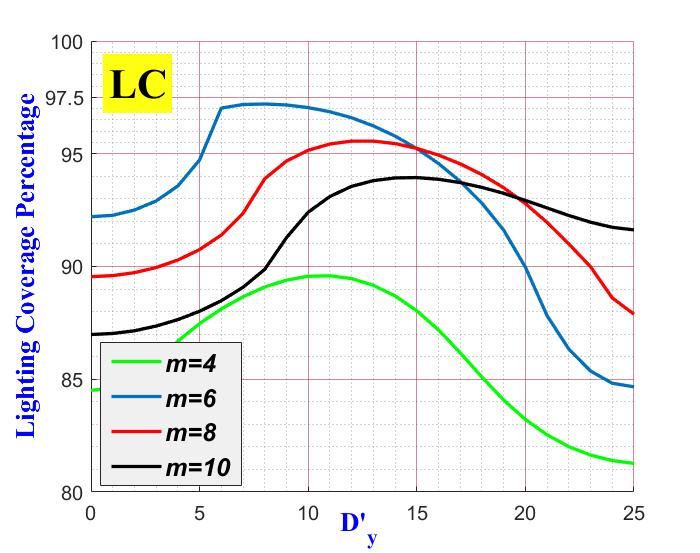}}%
		\subfloat[ \label{fig5_3b}]{\includegraphics[width=0.25\linewidth,height=0.14\textheight]{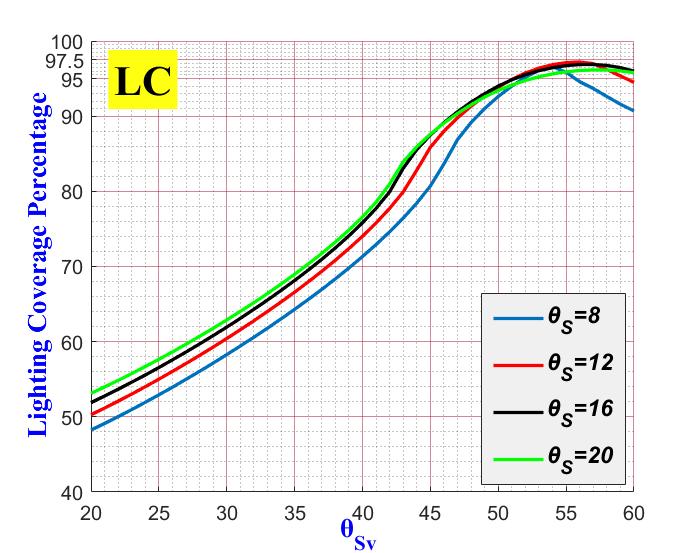}}%
		\subfloat[ \label{fig5_3c}]{\includegraphics[width=0.25\linewidth,height=0.14\textheight]{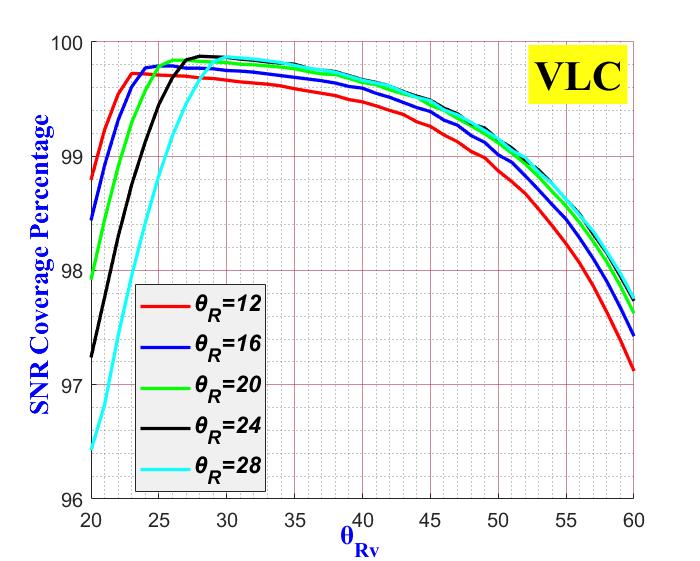}}%
		\subfloat[ \label{fig5_3d}]{\includegraphics[width=0.25\linewidth,height=0.14\textheight]{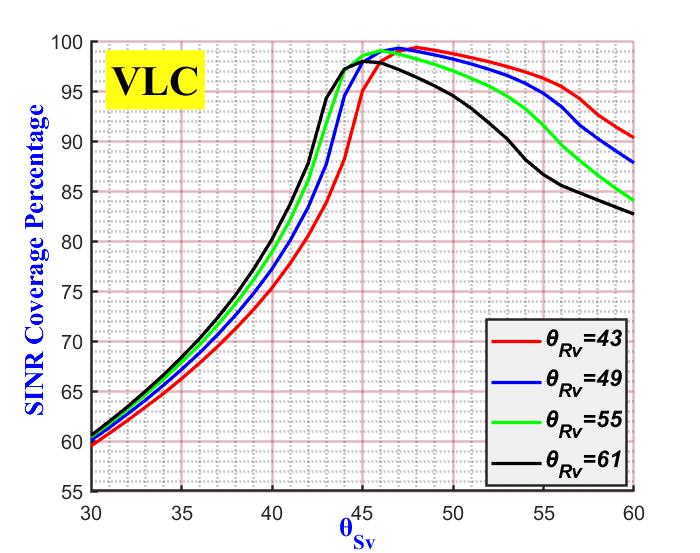}}       
		\vspace{-0.2cm}
		\subfloat[ \label{fig5_3e}]{\includegraphics[width=0.25\linewidth,height=0.14\textheight]{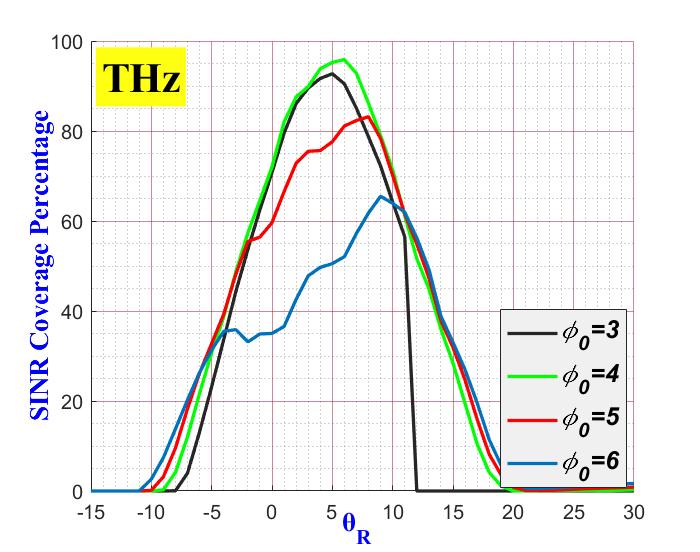}}%
		\subfloat[ \label{fig5_3f}]{\includegraphics[width=0.25\linewidth,height=0.14\textheight]{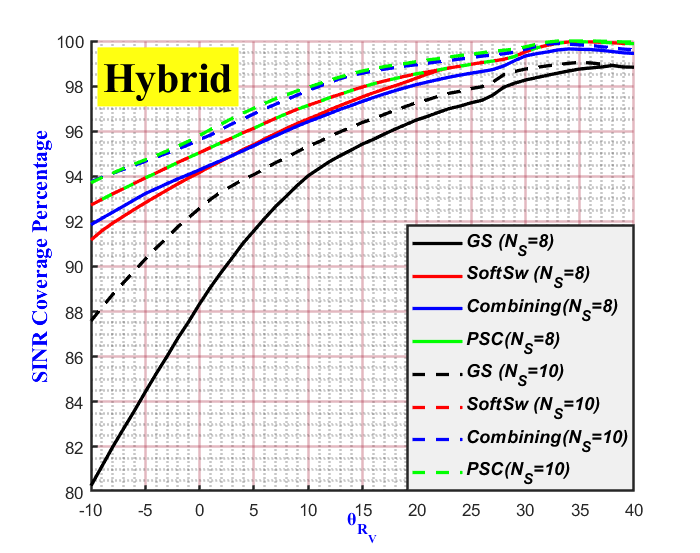}}%
		\subfloat[ \label{fig5_3g}]{\includegraphics[width=0.25\linewidth,height=0.14\textheight]{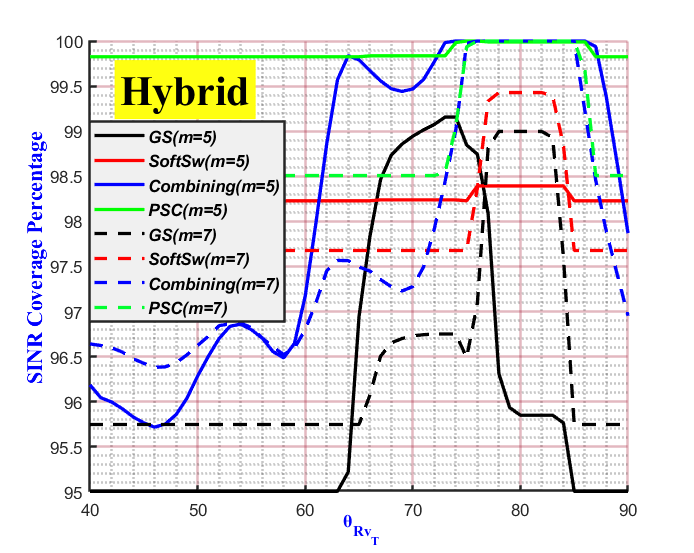}}%
		\subfloat[ \label{fig5_3h}]{\includegraphics[width=0.25\linewidth,height=0.14\textheight]{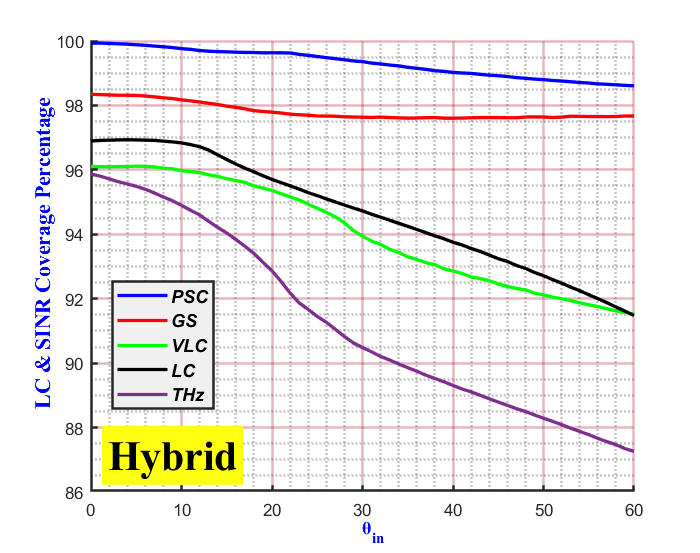}}%
		\caption{Comparative graphs of the coverage percentage according to (a)lighting in different values of m and $D'_y$, (b)lighting in different values of $\theta_S$ and $\theta_{Sv}$, (c)VLC SNR in different values of $\theta_R$ and $\theta_{Rv}$, (d)VLC SINR in different values of $\theta_{Rv}$ and $\theta_{Sv}$, (e)THz SINR in different values of $\phi_0$ and $\theta_R$, \textcolor{black}{(f)  SINR in different values of $N_R$ and $\theta_{R_V}$  for comparing the performance of different switching mechanisms, (g)  SINR in different values of $m$ and $\theta_{R_{v_T}}$ for comparing different switching  systems,} (h) LC and SINR in different values of $\theta_in $ for PSC, GS, VLC, and THz systems.}
        \vspace{-0.3cm}
		\label{fig:fig5_3}
	\end{figure*}
	
Subsequently, we focus on designing a backup THz system to complement our VLC system. As previously noted, the default values of the THz system parameters are provided in Table \ref{jad2}. Upon evaluating the SIC with these default values, we obtain Fig. \ref{fig5_1i}, which reveals a notably low coverage of $5.99\%$.  To address this, we optimize the THz system and derive optimal parameter values, which are detailed in the second column of Table \ref{jad4}. The results of this optimization are illustrated in Fig. \ref{fig5_1j}, where the SIC improves significantly to $95.87\%$. When we assess our hybrid VLC/THz system under these optimized conditions using the GS method, the SIC reaches $95.96\%$. Furthermore, with the implementation of the PSC method, the SIC is enhanced to $99.41\%$.

  As we mentioned, the major problem in VLC communications is related to weather conditions, which reduces the visibility capability. Also, background light from other light sources, especially sunlight during the day, intensifies this problem. In this case, the hybrid system bears the burden of this problem. Table \ref{jad5} presents some different scenarios of the effect of visibility  and background light power (BG power) along with the coverage of the VLC system and the hybrid system for SINR. In the scenario where visibility and light interference from other sources are at their worst, we apply an optimization step to the THz system parameters. We also apply a trade-off to the results to maintain the quality of the single THz system. In a way that  the coverage percentage in the non-optimized mode of PSC (NPSC) increases from $98.64\%$ to $99.12\%$. And for the corresponding GS, it also reaches $94.84\%$ considering that the VLC system has dropped drastically. The last column of Table \ref{jad4} and Fig \ref{fig5_2b} shows these selected values for the THz system, and Figs. \ref{fig5_2c} and \ref{fig5_2d} show the SIC for GS and PSC, respectively. As it is known, the superiority of the PSC system over the GS system is noticeable. Also, this system shows great resistance in different weather conditions, and at most of the conditions the coverage remains stable at more than $99.9\%$ as shown in Table \ref{jad5}.

  Finally, we aim to evaluate the outage probability coverage (OPC) on the street for the cases outlined in Table \ref{jad4}. The results are presented in Figs. \ref{fig5_2e}, \ref{fig5_2f}, \ref{fig5_2g}, and \ref{fig5_2h}, which depict the OPC for the VLC system, the THz system, and the hybrid system under both the GS and PSC scenarios, respectively.
As evident from the figures, the hybrid system offers superior coverage compared to the VLC and THz systems operating independently. Additionally, our proposed PSC method continues to outperform the GS method in terms of OPC, further confirming its advantage in maintaining reliable communication across the street surface.

 The content presented in Fig. \ref{fig:fig5_3} confirms the optimal parameter values obtained. We have included a selection of graphs derived from the results. As shown in Fig. \ref{fig5_3a}, the percentage of LC  versus the $ D_y^{\prime} $ is depicted for four different values of $ m $, indicating that the optimal condition is achieved for a distance of 8 meters at $ m=6 $. Similarly, in Fig. \ref{fig5_3b}, it's evident that the obtained value of $\theta_{S_v}$, namely $56^\circ$, is the optimal value for LC.  Figure \ref{fig5_3c} illustrates an example validating the selected parameter values for SC. In contrast, Figures \ref{fig5_3d} and \ref{fig5_3e} focus on the SIC of the VLC and THz systems, respectively. \textcolor{black}{The subsequent  two graphs illustrate the SIC performance of the hybrid system, highlighting its behavior across various switching schemes: GS, Soft Sw, Combining, and PSC. It is evident that our PSC system not only incorporates multiple advantages but also outperforms the others in terms of link quality and stability of the communication system. Finally, Fig. \ref{fig5_3h} presents a comparison of the hybrid VLC/THz system’s performance against that of the individual VLC and THz systems, all evaluated under the same input angle parameter, $\theta_{in}$.}

\section{\textcolor{black}{Conclusion and Future Directions}}

\textcolor{black}{This paper presented a comprehensive framework for an ITS that simultaneously supports optimal street lighting and robust I2V communication. Building on our earlier work in \cite{ref111}, which focused on the backhaul network, we modeled and optimized the access network using a hybrid VLC/THz architecture. By applying a grid search optimization approach, we systematically evaluated thousands of configurations to maximize lighting coverage (LC), received power, SNR, and SINR. The results demonstrate significant performance gains: lighting coverage increased from 35\% to 97\% and hybrid communication coverage improved from 49\% to 99.9\% at a given power level. Under highly adverse conditions, the hybrid system further boosted coverage from 69\% (VLC-only) to 99\%, ensuring continuous connectivity. Optimized THz parameters, combined with the PSC strategy, achieved an impressive 99.94\% communication coverage. Outage probability analysis confirmed the system’s robustness. These improvements highlight the flexibility, efficiency, and scalability of our proposed architecture as a foundation for future ITS deployments.}

\textcolor{black}{With these achievements, we have completed the conceptual and structural design of a future-ready smart traffic system. This framework is the result of carefully integrating state-of-the-art solutions from the scientific literature into a unified, high-performance architecture. Key innovations include the adoption of cell-free networking, the use of FSO, VLC, and THz communications, advanced physical design strategies, and the development of a dynamic switching-combining mechanism. These elements collectively form the backbone of our proposed model. To ensure realistic performance evaluation, the system’s parameters and configuration have been aligned as closely as possible with experimental and real-world conditions. Furthermore, by incorporating actual urban standards and constraints into the physical design of network components, we have taken a significant step toward practical implementation. Despite these advancements, further refinement is necessary to transition   to real-world deployment. Future work will focus on the phase of detailed communication system design, equipment-level configuration, and field validation. In addition, the modular innovations across multiple layers of the proposed architecture offer a strong foundation for future research, development, and real-world application. We hope that this work makes a meaningful contribution to the ongoing evolution of ITS and smart city communication systems, and provides a valuable reference for both researchers and engineers in the field.}

	\bibliographystyle{IEEEtranN}
	{\small
		\bibliography{references}}

\begin{thebibliography}{54}
\providecommand{\natexlab}[1]{#1}
\providecommand{\url}[1]{#1}
\csname url@samestyle\endcsname
\providecommand{\newblock}{\relax}
\providecommand{\bibinfo}[2]{#2}
\providecommand{\BIBentrySTDinterwordspacing}{\spaceskip=0pt\relax}
\providecommand{\BIBentryALTinterwordstretchfactor}{4}
\providecommand{\BIBentryALTinterwordspacing}{\spaceskip=\fontdimen2\font plus
\BIBentryALTinterwordstretchfactor\fontdimen3\font minus
  \fontdimen4\font\relax}
\providecommand{\BIBforeignlanguage}[2]{{%
\expandafter\ifx\csname l@#1\endcsname\relax
\typeout{** WARNING: IEEEtranN.bst: No hyphenation pattern has been}%
\typeout{** loaded for the language `#1'. Using the pattern for}%
\typeout{** the default language instead.}%
\else
\language=\csname l@#1\endcsname
\fi
#2}}
\providecommand{\BIBdecl}{\relax}
\BIBdecl

\bibitem[Abdel~Hakeem et~al.(2022)Abdel~Hakeem, Hussein, and Kim]{ref3}
S.~A. Abdel~Hakeem, H.~H. Hussein, and H.~Kim, ``Security requirements and
  challenges of 6g technologies and applications,'' \emph{Sensors}, vol.~22,
  no.~5, p. 1969, 2022.

\bibitem[Murroni et~al.(2023)Murroni, Anedda, Fadda, Ruiu, Popescu, Zaharia,
  and Giusto]{ref4}
M.~Murroni, M.~Anedda, M.~Fadda, P.~Ruiu, V.~Popescu, C.~Zaharia, and
  D.~Giusto, ``6g—enabling the new smart city: A survey,'' \emph{Sensors},
  vol.~23, no.~17, p. 7528, 2023.

\bibitem[Meucci et~al.(2021)Meucci, Seminara, Nawaz, Caputo, Mucchi, and
  Catani]{ref28}
M.~Meucci, M.~Seminara, T.~Nawaz, S.~Caputo, L.~Mucchi, and J.~Catani,
  ``Bidirectional vehicle-to-vehicle communication system based on vlc: Outdoor
  tests and performance analysis,'' \emph{IEEE Transactions on Intelligent
  Transportation Systems}, vol.~23, no.~8, pp. 11\,465--11\,475, 2021.

\bibitem[Eldeeb et~al.(2022)Eldeeb, Elamassie, Sait, and Uysal]{ref29}
H.~B. Eldeeb, M.~Elamassie, S.~M. Sait, and M.~Uysal,
  ``Infrastructure-to-vehicle visible light communications: Channel modelling
  and performance analysis,'' \emph{IEEE Transactions on Vehicular Technology},
  vol.~71, no.~3, pp. 2240--2250, 2022.

\bibitem[Patel et~al.(2021)Patel, Shah, Ding, Guan, Sun, Chang, and Lim]{ref30}
D.~K. Patel, H.~Shah, Z.~Ding, Y.~L. Guan, S.~Sun, Y.~C. Chang, and J.~M.-Y.
  Lim, ``Performance analysis of noma in vehicular communications over inid
  nakagami-m fading channels,'' \emph{IEEE Transactions on Wireless
  Communications}, vol.~20, no.~10, pp. 6254--6268, 2021.

\bibitem[Liu et~al.(2020)Liu, Sarfraz, and Wang]{ref12}
Q.~Liu, S.~Sarfraz, and S.~Wang, ``An overview of key technologies and
  challenges of 6g,'' in \emph{Machine Learning for Cyber Security: Third
  International Conference, ML4CS 2020, Guangzhou, China, October 8--10, 2020,
  Proceedings, Part II 3}.\hskip 1em plus 0.5em minus 0.4em\relax Springer,
  2020, pp. 315--326.

\bibitem[Marabissi et~al.(2020)Marabissi, Mucchi, Caputo, Nizzi, Pecorella,
  Fantacci, Nawaz, Seminara, and Catani]{ref19}
D.~Marabissi, L.~Mucchi, S.~Caputo, F.~Nizzi, T.~Pecorella, R.~Fantacci,
  T.~Nawaz, M.~Seminara, and J.~Catani, ``Experimental measurements of a joint
  5g-vlc communication for future vehicular networks,'' \emph{Journal of Sensor
  and Actuator Networks}, vol.~9, no.~3, p.~32, 2020.

\bibitem[Ibhaze et~al.(2020)Ibhaze, Orukpe, and Edeko]{ref21}
A.~E. Ibhaze, P.~E. Orukpe, and F.~O. Edeko, ``High capacity data rate system:
  Review of visible light communications technology,'' \emph{Journal of
  Electronic Science and Technology}, vol.~18, no.~3, p. 100055, 2020.

\bibitem[Yang et~al.(2020)Yang, Zhong, Chen, and Alphones]{ref22}
H.~Yang, W.-D. Zhong, C.~Chen, and A.~Alphones, ``Integration of visible light
  communication and positioning within 5g networks for internet of things,''
  \emph{IEEE Network}, vol.~34, no.~5, pp. 134--140, 2020.

\bibitem[Kamruzzaman(2022)]{ref13}
M.~Kamruzzaman, ``Key technologies, applications and trends of internet of
  things for energy-efficient 6g wireless communication in smart cities,''
  \emph{Energies}, vol.~15, no.~15, p. 5608, 2022.

\bibitem[Mar{\`e} et~al.(2016)Mar{\`e}, Cugnasca, Marte, and Gentile]{ref13_1}
R.~M. Mar{\`e}, C.~E. Cugnasca, C.~L. Marte, and G.~Gentile, ``Intelligent
  transport systems and visible light communication applications: An
  overview,'' in \emph{2016 IEEE 19th International Conference on Intelligent
  Transportation Systems (ITSC)}.\hskip 1em plus 0.5em minus 0.4em\relax IEEE,
  2016, pp. 2101--2106.

\bibitem[Borogovac et~al.(2011)Borogovac, Rahaim, Tuganbayeva, and
  Little]{ref48}
T.~Borogovac, M.~B. Rahaim, M.~Tuganbayeva, and T.~D. Little,
  ``“lights-off” visible light communications,'' in \emph{2011 IEEE
  GLOBECOM Workshops (GC Wkshps)}.\hskip 1em plus 0.5em minus 0.4em\relax IEEE,
  2011, pp. 797--801.

\bibitem[Rajahrajasingh and Jayakody(2024)]{ref114}
H.~Rajahrajasingh and D.~N.~K. Jayakody, ``Unmanned aerial vehicle-assisted
  terahertz--visible light communication systems: An in-depth performance
  analysis,'' \emph{Sensors}, vol.~24, no.~13, p. 4080, 2024.

\bibitem[Singh et~al.(2024)Singh, Srivastava, Bohara, Noor-A-Rahim, Liu, and
  Pesch]{ref115}
G.~Singh, A.~Srivastava, V.~A. Bohara, M.~Noor-A-Rahim, Z.~Liu, and D.~Pesch,
  ``Toward 6g-v2x: Aggregated rf-vlc for ultra-reliable and low-latency
  autonomous driving,'' \emph{IEEE Communications Standards Magazine}, vol.~8,
  no.~4, pp. 80--87, 2024.

\bibitem[Demir et~al.(2020)Demir, Eldeeb, and Uysal]{ref77}
M.~S. Demir, H.~B. Eldeeb, and M.~Uysal, ``Comp-based dynamic handover for
  vehicular vlc networks,'' \emph{IEEE Communications Letters}, vol.~24, no.~9,
  pp. 2024--2028, 2020.

\bibitem[Eldeeb et~al.(2021{\natexlab{a}})Eldeeb, Sait, and Uysal]{ref73}
H.~B. Eldeeb, S.~M. Sait, and M.~Uysal, ``Visible light communication for
  connected vehicles: How to achieve the omnidirectional coverage?'' \emph{IEEE
  Access}, vol.~9, pp. 103\,885--103\,905, 2021.

\bibitem[Eldeeb et~al.(2021{\natexlab{b}})Eldeeb, Elamassie, and Uysal]{ref83}
H.~B. Eldeeb, M.~Elamassie, and M.~Uysal, ``Performance analysis and
  optimization of cascaded i2v and v2v vlc links,'' in \emph{2021 17th
  International Symposium on Wireless Communication Systems (ISWCS)}.\hskip 1em
  plus 0.5em minus 0.4em\relax IEEE, 2021, pp. 1--6.

\bibitem[Abouzohri and Abdallah(2020)]{ref49}
E.~M.~H. Abouzohri and M.~M. Abdallah, ``Performance of hybrid cognitive rf/vlc
  systems in vehicle-to-vehicle communications,'' in \emph{2020 IEEE
  International Conference on Informatics, IoT, and Enabling Technologies
  (ICIoT)}.\hskip 1em plus 0.5em minus 0.4em\relax IEEE, 2020, pp. 429--434.

\bibitem[Singya et~al.(2022)Singya, Makki, D’Errico, and Alouini]{ref62}
P.~K. Singya, B.~Makki, A.~D’Errico, and M.-S. Alouini, ``Hybrid
  fso/thz-based backhaul network for mmwave terrestrial communication,''
  \emph{IEEE Transactions on Wireless Communications}, 2022.

\bibitem[Singya et~al.(2023)Singya, Makki, D'Errico, and Alouini]{ref56}
P.~K. Singya, B.~Makki, A.~D'Errico, and M.-S. Alouini, ``High-rate reliable
  communication using multi-hop and mesh thz/fso networks,'' \emph{arXiv
  preprint arXiv:2304.01643}, 2023.

\bibitem[Eldeeb and Uysal(2019)]{ref74}
H.~B. Eldeeb and M.~Uysal, ``Vehicle-to-vehicle visible light communication:
  How to select receiver locations for optimal performance?'' in \emph{2019
  11th International Conference on Electrical and Electronics Engineering
  (ELECO)}.\hskip 1em plus 0.5em minus 0.4em\relax IEEE, 2019, pp. 402--405.

\bibitem[Seminara et~al.(2020)Seminara, Nawaz, Caputo, Mucchi, and
  Catani]{ref75}
M.~Seminara, T.~Nawaz, S.~Caputo, L.~Mucchi, and J.~Catani, ``Characterization
  of field of view in visible light communication systems for intelligent
  transportation systems,'' \emph{IEEE Photonics Journal}, vol.~12, no.~4, pp.
  1--16, 2020.

\bibitem[You et~al.(2020)You, Zhong, Chen, and Yu]{ref76}
X.~You, Y.~Zhong, J.~Chen, and C.~Yu, ``Mobile channel estimation based on
  decision feedback in vehicle-to-infrastructure visible light communication
  systems,'' \emph{Optics Communications}, vol. 462, p. 125261, 2020.

\bibitem[Abdalla and Cooper(2023)]{ref55}
R.~Abdalla and A.~B. Cooper, ``Performance analysis of los thz systems under
  misalignment and deterministic fading,'' in \emph{2023 57th Annual Conference
  on Information Sciences and Systems (CISS)}.\hskip 1em plus 0.5em minus
  0.4em\relax IEEE, 2023, pp. 1--5.

\bibitem[Li and Yang(2021)]{ref57}
S.~Li and L.~Yang, ``Performance analysis of dual-hop thz transmission systems
  over $\alpha$-$\mu$ fading channels with pointing errors,'' \emph{IEEE
  Internet of Things Journal}, vol.~9, no.~14, pp. 11\,772--11\,783, 2021.

\bibitem[Teli et~al.(2018)Teli, Zvanovec, and Ghassemlooy]{ref16}
S.~R. Teli, S.~Zvanovec, and Z.~Ghassemlooy, ``Optical internet of things
  within 5g: Applications and challenges,'' in \emph{2018 IEEE International
  Conference on Internet of Things and Intelligence System (IOTAIS)}.\hskip 1em
  plus 0.5em minus 0.4em\relax IEEE, 2018, pp. 40--45.

\bibitem[Eldeeb et~al.(2018)Eldeeb, Selmy, Elsayed, and Badr]{ref42}
H.~B. Eldeeb, H.~A. Selmy, H.~M. Elsayed, and R.~I. Badr, ``Interference
  mitigation and capacity enhancement using constraint field of view adr in
  downlink vlc channel,'' \emph{IET Communications}, vol.~12, no.~16, pp.
  1968--1978, 2018.

\bibitem[Madani et~al.(2017)Madani, Baghersalimi, and Ghassemlooy]{ref46}
F.~Madani, G.~Baghersalimi, and Z.~Ghassemlooy, ``Effect of transmitter and
  receiver parameters on the output signal to noise ratio in visible light
  communications,'' in \emph{2017 Iranian Conference on Electrical Engineering
  (ICEE)}.\hskip 1em plus 0.5em minus 0.4em\relax IEEE, 2017, pp. 2111--2116.

\bibitem[Chaleshtori et~al.(2020)Chaleshtori, Zvanovec, Ghassemlooy, Eldeeb,
  and Uysal]{ref67}
Z.~N. Chaleshtori, S.~Zvanovec, Z.~Ghassemlooy, H.~B. Eldeeb, and M.~Uysal,
  ``Coverage of a shopping mall with flexible oled-based visible light
  communications,'' \emph{Optics express}, vol.~28, no.~7, pp.
  10\,015--10\,026, 2020.

\bibitem[Eso et~al.(2021)Eso, Ghassemlooy, Zvanovec, Sathian, and
  Gholami]{ref69}
E.~Eso, Z.~Ghassemlooy, S.~Zvanovec, J.~Sathian, and A.~Gholami, ``Fundamental
  analysis of vehicular light communications and the mitigation of sunlight
  noise,'' \emph{IEEE Transactions on Vehicular Technology}, vol.~70, no.~6,
  pp. 5932--5943, 2021.

\bibitem[Ghassemlooy et~al.(2019)Ghassemlooy, Popoola, and Rajbhandari]{ref71}
Z.~Ghassemlooy, W.~Popoola, and S.~Rajbhandari, \emph{Optical wireless
  communications: system and channel modelling with
  Matlab{\textregistered}}.\hskip 1em plus 0.5em minus 0.4em\relax CRC press,
  2019.

\bibitem[Eldeeb et~al.(2021{\natexlab{c}})Eldeeb, Eso, Jarchlo, Zvanovec,
  Uysal, Ghassemlooy, and Sathian]{ref68}
H.~B. Eldeeb, E.~Eso, E.~A. Jarchlo, S.~Zvanovec, M.~Uysal, Z.~Ghassemlooy, and
  J.~Sathian, ``Vehicular vlc: A ray tracing study based on measured radiation
  patterns of commercial taillights,'' \emph{IEEE Photonics Technology
  Letters}, vol.~33, no.~16, pp. 904--907, 2021.

\bibitem[Karbalayghareh et~al.(2020)Karbalayghareh, Miramirkhani, Eldeeb,
  Kizilirmak, Sait, and Uysal]{ref70}
M.~Karbalayghareh, F.~Miramirkhani, H.~B. Eldeeb, R.~C. Kizilirmak, S.~M. Sait,
  and M.~Uysal, ``Channel modelling and performance limits of vehicular visible
  light communication systems,'' \emph{IEEE Transactions on Vehicular
  Technology}, vol.~69, no.~7, pp. 6891--6901, 2020.

\bibitem[Aly et~al.(2021{\natexlab{a}})Aly, Elamassie, and Uysal]{ref82}
B.~Aly, M.~Elamassie, and M.~Uysal, ``Vehicular vlc channel model for a
  low-beam headlight transmitter,'' in \emph{2021 17th International Symposium
  on Wireless Communication Systems (ISWCS)}.\hskip 1em plus 0.5em minus
  0.4em\relax IEEE, 2021, pp. 1--5.

\bibitem[Turan et~al.(2022)Turan, Narmanlioglu, Koc, Kar, Coleri, and
  Uysal]{ref80}
B.~Turan, O.~Narmanlioglu, O.~N. Koc, E.~Kar, S.~Coleri, and M.~Uysal,
  ``Measurement based non-line-of-sight vehicular visible light communication
  channel characterization,'' \emph{IEEE Transactions on Vehicular Technology},
  vol.~71, no.~9, pp. 10\,110--10\,114, 2022.

\bibitem[Aly et~al.(2021{\natexlab{b}})Aly, Elamassie, and Uysal]{ref81}
B.~Aly, M.~Elamassie, and M.~Uysal, ``Experimental characterization of
  multi-hop vehicular vlc systems,'' in \emph{2021 IEEE 32nd Annual
  International Symposium on Personal, Indoor and Mobile Radio Communications
  (PIMRC)}.\hskip 1em plus 0.5em minus 0.4em\relax IEEE, 2021, pp. 1--6.

\bibitem[Vieira et~al.(2020)Vieira, Vieira, Louro, and Vieira]{ref78}
M.~A. Vieira, M.~Vieira, P.~Louro, and P.~Vieira, ``Redesign of the trajectory
  within a complex intersection for visible light communication ready connected
  cars,'' \emph{Optical Engineering}, vol.~59, no.~9, pp. 097\,104--097\,104,
  2020.

\bibitem[Vieira et~al.(2018)Vieira, Vieira, Louro, and Vieira]{ref79}
------, ``Cooperative vehicular communication systems based on visible light
  communication,'' \emph{Optical Engineering}, vol.~57, no.~7, pp.
  076\,101--076\,101, 2018.

\bibitem[Huang et~al.(2023)Huang, Zheng, and Zhang]{ref85}
Z.~Huang, B.~Zheng, and R.~Zhang, ``Roadside irs-aided vehicular communication:
  Efficient channel estimation and low-complexity beamforming design,''
  \emph{IEEE Transactions on Wireless Communications}, 2023.

\bibitem[Han et~al.(2018)Han, Zhou, Yang, and Li]{ref90}
Y.~Han, X.~Zhou, L.~Yang, and S.~Li, ``A bipartite matching based user pairing
  scheme for hybrid vlc-rf noma systems,'' in \emph{2018 International
  Conference on Computing, Networking and Communications (ICNC)}.\hskip 1em
  plus 0.5em minus 0.4em\relax IEEE, 2018, pp. 480--485.

\bibitem[Pan et~al.(2019)Pan, Lei, Ding, and Ni]{ref92}
G.~Pan, H.~Lei, Z.~Ding, and Q.~Ni, ``3-d hybrid vlc-rf indoor iot systems with
  light energy harvesting,'' \emph{IEEE Transactions on Green Communications
  and Networking}, vol.~3, no.~3, pp. 853--865, 2019.

\bibitem[Modami et~al.(2024)Modami, Beiranvand, and Dabiri]{ref111}
Y.~Modami, H.~Beiranvand, and M.~T. Dabiri, ``Cell-free
  infrastructure-to-vehicle communication: Empowering with hybrid fso/thz
  backhaul,'' \emph{IEEE Wireless Communications Letters}, 2024.

\bibitem[Hegr et~al.(2015)Hegr, Voznak, Kozak, and Bohac]{ref118}
T.~Hegr, M.~Voznak, M.~Kozak, and L.~Bohac, ``Measurement of switching latency
  in high data rate ethernet networks,'' \emph{Elektronika ir Elektrotechnika},
  vol.~21, no.~3, pp. 73--78, 2015.

\bibitem[3GPP(2018)]{ref109}
3GPP, ``Study on channel model for frequencies from 0.5 to 100 ghz,'' 2018.

\bibitem[Yahia et~al.(2021)Yahia, Meraihi, Ramdane-Cherif, Gabis, Acheli, and
  Guan]{ref40}
S.~Yahia, Y.~Meraihi, A.~Ramdane-Cherif, A.~B. Gabis, D.~Acheli, and H.~Guan,
  ``A survey of channel modeling techniques for visible light communications,''
  \emph{Journal of Network and Computer Applications}, vol. 194, p. 103206,
  2021.

\bibitem[Zaki et~al.(2019)Zaki, Fayed, Abd El~Aziz, and Aly]{ref102}
R.~W. Zaki, H.~A. Fayed, A.~Abd El~Aziz, and M.~H. Aly, ``Outdoor visible light
  communication in intelligent transportation systems: Impact of snow and
  rain,'' \emph{Applied Sciences}, vol.~9, no.~24, p. 5453, 2019.

\bibitem[Abualhoul(2016)]{ref103}
M.~Abualhoul, ``Visible light and radio communication for cooperative
  autonomous driving: applied to vehicle convoy,'' Ph.D. dissertation,
  Universit{\'e} Paris sciences et lettres, 2016.

\bibitem[Balanis(2016)]{ref112}
C.~A. Balanis, \emph{Antenna theory: analysis and design}.\hskip 1em plus 0.5em
  minus 0.4em\relax John wiley \& sons, 2016.

\bibitem[Sharma et~al.(2019)Sharma, Madhukumar, and Swaminathan]{ref58}
S.~Sharma, A.~Madhukumar, and R.~Swaminathan, ``Effect of pointing errors on
  the performance of hybrid fso/rf networks,'' \emph{IEEE Access}, vol.~7, pp.
  131\,418--131\,434, 2019.

\bibitem[Dabiri et~al.(2023)Dabiri, Hasna, and Khattab]{ref60}
M.~T. Dabiri, M.~Hasna, and T.~Khattab, ``Thz vs. fso: An outage probability
  and channel capacity performance comparison study,'' in \emph{2023
  International Symposium on Networks, Computers and Communications
  (ISNCC)}.\hskip 1em plus 0.5em minus 0.4em\relax IEEE, 2023, pp. 1--6.

\bibitem[Papasotiriou et~al.(2021{\natexlab{a}})Papasotiriou, Boulogeorgos,
  Haneda, de~Guzman, and Alexiou]{ref106}
E.~N. Papasotiriou, A.-A.~A. Boulogeorgos, K.~Haneda, M.~F. de~Guzman, and
  A.~Alexiou, ``An experimentally validated fading model for thz wireless
  systems,'' \emph{Scientific Reports}, vol.~11, no.~1, p. 18717, 2021.

\bibitem[Papasotiriou et~al.(2021{\natexlab{b}})Papasotiriou, Boulogeorgos, and
  Alexiou]{ref107}
E.~N. Papasotiriou, A.-A.~A. Boulogeorgos, and A.~Alexiou, ``fading modeling in
  indoor thz wireless systems,'' in \emph{2021 International Balkan Conference
  on Communications and Networking (BalkanCom)}.\hskip 1em plus 0.5em minus
  0.4em\relax IEEE, 2021, pp. 161--165.

\bibitem[Dabiri et~al.(2024)Dabiri, Althunibat, Hasna, and Qaraqe]{ref113}
M.~T. Dabiri, S.~Althunibat, M.~Hasna, and K.~Qaraqe, ``Non-orthogonal multiple
  access scheme using directional thz antennas under positioning errors,''
  \emph{IEEE Transactions on Vehicular Technology}, 2024.

\bibitem[Liu et~al.(2012)Liu, Chan, Ng, Lo, and Shimamoto]{ref108}
J.~Liu, P.~W.~C. Chan, D.~W.~K. Ng, E.~S. Lo, and S.~Shimamoto, ``Hybrid
  visible light communications in intelligent transportation systems with
  position based services,'' in \emph{2012 IEEE Globecom Workshops}.\hskip 1em
  plus 0.5em minus 0.4em\relax IEEE, 2012, pp. 1254--1259.

\end{thebibliography}

	\vfill
	
\end{document}